\documentclass[aps,pra,twocolumn, 10pt, showpacs,superscriptaddress,groupedaddress]{revtex4-1}  

\newcommand{\braket}[1]{\ensuremath{\left< #1 \right>}}

\newcommand{\Rb}{\ensuremath{^{87}\text{Rb }}}

\newcommand{\pd}[2]{\ensuremath{\frac{\partial #1}{\partial #2}}}

\newcommand{\ie}{\emph{i.e.} }

\newcommand{\Ha}{\mathcal{H}}

\usepackage{graphicx}
\usepackage{epstopdf}
\usepackage{setspace}
\usepackage{bbold}
\usepackage{comment}
\usepackage{color}
\usepackage{soul}
\usepackage[abs]{overpic}
\usepackage{amsmath}
\usepackage{lipsum}
\usepackage[abs]{overpic}
\usepackage{amsmath}
\usepackage{nicefrac}

\begin{document}

\author{Gadi Afek}
\altaffiliation[Present address: ]{Wright Laboratory, Department of Physics, Yale University, New Haven, Connecticut 06520, USA}\affiliation{Department of Physics of Complex Systems, Weizmann Institute of Science, Rehovot 76100, Israel }
\author{Alexander Cheplev}
\affiliation{Department of Physics of Complex Systems, Weizmann Institute of Science, Rehovot 76100, Israel }
\author{Arnaud Courvoisier}
\affiliation{Department of Physics of Complex Systems, Weizmann Institute of Science, Rehovot 76100, Israel }
\author{Nir Davidson}
\affiliation{Department of Physics of Complex Systems, Weizmann Institute of Science, Rehovot 76100, Israel }

\title[]{Deviations from generalized Equipartition in Confined, Laser Cooled Atoms}

\begin{abstract}
 We observe a significant steady-state deviation from the generalized equipartition theorem, one of the pivotal results of classical statistical mechanics, in a system of confined, laser cooled atoms. We quantify this deviation, measure its dynamics and show that its steady state value quantifies the departure of non-thermal states from thermal equilibrium even for anharmonic confinement. In particular, we find that deviations from equipartition grow as the system dynamics becomes more anomalous. We present numerical simulations that validate the experimental data and reveal an inhomogeneous distribution of the kinetic energy through the system, supported by an analytical analysis of the phase space.
\end{abstract}

\maketitle

The 100 year-old generalized equipartition theorem~\cite{tolman1918general}, one of the cornerstones of classical statistical physics, remains of great interest throughout various fields of research to this day~\cite{Uline2008,Maggi2014,Chen2017}. It states that for a system in thermal equilibrium with a heat bath of temperature $T$, for any generalized coordinate $q_i$ and Hamiltonian $\Ha$:
\begin{equation}\label{eq:equipartition}
    \braket{q_i\pd{\Ha}{q_j}}=\delta_{ij}k_BT
\end{equation}
where $k_B$ is the Boltzmann constant, $\delta_{ij}$ is a Kronecker Delta and $\braket{...}$ denotes ensemble averaging. An immediate result of this theorem is the well known equipartition theorem~\cite{Waterston1892,Boltzmann1877}, valid for degrees of freedom which appear quadratically in the Hamiltonian, in which case the relation implies equipartition of energy among those degrees of freedom. Other significant extensions have been shown for finite sized systems~\cite{Mello2010}, generalized Canonical ensembles~\cite{Frisch53} and non-extensive thermodynamics~\cite{Martinez02}. While the equipartition theorem is strictly correct only in thermal equilibrium, it was extended and applied to other thermodynamic systems and observed to hold even outside of thermal equilibrium~\cite{Anderson1995}.

Ultracold atomic systems have been pushing the understanding of statistical physics for several decades, and have recently began to explore aspects of non-equilibrium physics~\cite{Diehl2008,Diehl2011,Lobser2015,Kindermann2016Nature,Mayer2019}. An especially interesting system for probing non-equilibrium statistical mechanics is that of ultracold atomic ensembles in dissipative one-dimensional (1D) optical lattices, where the heat bath is implemented by the field of the lattice lasers. The main advantage of such a system is the unique control over experimental parameters, allowing fine-tuning of the dynamics. In addition to it being an experimentally and theoretically well established test bed for anomalous dynamics~\cite{Castin1991,Marksteiner1996,Katori1997,Lutz2003,Lutz2004,Jersblad2004,Douglas2006,Kessler2010,Hirschberg2011,Sagi2012,Kessler2012,Dechant2012,Wickenbrock2012,Barkai2014,Dechant2014,Zaburdaev2015,Holz2015,Afek2017Correlations}, it has recently been linked with the notion of non-thermal equilibrium~\cite{Dechant2015,Dechant2016}. Through extensive analysis of the phase space dynamics of such a system confined in a harmonic potential, a prediction has been put forth of a violation, under certain conditions, of the equipartition theorem. 

In this Paper, we present a detailed experimental investigation of the steady-state deviation from equipartition for trapped, laser cooled atoms in contact with a non-thermal heat bath, implemented by a 1D, dissipative optical lattice. For completion, we also investigate numerically the effect of anharmonicity of the confining potential on the dynamics and magnitude of this deviation. Finally, as a basis for further work, we present a novel prediction of the position dependence of the local kinetic energy for such confined Sisyphus cooled atoms, supported by analytics and numerics.

\subsection*{Using the equipartition theorem to quantify departure from thermal equilibrium}

Any departure of a 1D confined system with coordinates $(x,p)$ and Hamiltonian $\Ha$ from thermal equilibrium can be parametrized using the \textit{Equipartition parameter} $\chi$~\cite{Dechant2015,Dechant2016}:
\begin{equation}\label{eq:chi}
    \chi \equiv \sqrt{\left\langle p \frac{\partial \Ha}{\partial p}\right\rangle\bigg/\left\langle x \frac{\partial \Ha}{\partial x} \right\rangle}
\end{equation}
which depends explicitly on the details of the confining potential. Given that $\chi=1$ for systems in thermal equilibrium, deviations of $\chi$ from unity imply non-thermal distributions and a possible break of energy-probability equivalence. For the simple case of harmonic confinement, one can derive the harmonic equipartition parameter $\chi_H$ from Eq.~\ref{eq:chi},
\begin{equation}\label{eq:chi_definition}
    \chi_H \equiv \frac{\sigma_v}{\omega\sigma_x},
\end{equation}
where $\sigma_x$ and $\sigma_v$ are the respective standard deviations of the position and velocity distributions (both of which are Gaussian for a harmonic potential and thermal equilibrium), and $\omega$ is the harmonic trap frequency. $\chi_H$ is a leading order approximation of $\chi$ for any continuous potential with a minimum, and is, in practice, considerably easier to access experimentally than $\chi$ for many systems. 

\subsection*{Experimental}

\begin{figure} 
    \centering
        \begin{overpic}
            [width=\linewidth]{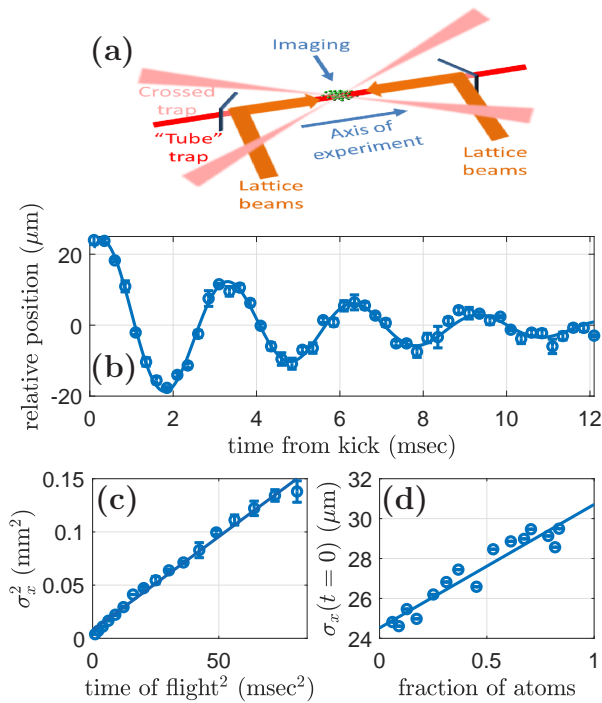}
            \put(32,250){\large \textbf{(a)}}
            \put(32,130){\large \textbf{(b)}}
            \put(34,80){\large \textbf{(c)}}
            \put(142,80){\large \textbf{(d)}}
        \end{overpic}
        \caption[Measuring equipartition in thermal equilibrium.]{Measurement scheme for the equipartition parameter, $\chi_H\equiv\sigma_v/\omega\sigma_x$. (a) A sketch of the experimental setup. Laser cooled \Rb atoms are trapped in a crossed dipole trap (light red) overlapped with a strong, single beam tube trap (bold red), and coupled to a non-thermal heat bath implemented by a set of 1D Sisyphus cooling lattice beams (orange) through dichroic mirrors. The atoms propagate in this combined potential and are subsequently imaged. (b) The trapped cloud is kicked with a short, directional pulse of near-resonant light to excite subsequent center of mass oscillations, and the center of mass position is extracted (circles). Trap frequency $\omega$ is calculated by fitting the data to an exponentially decaying sine. The decay of the center of mass oscillations is attributed to anharmonicity of the confining potential. (c) Typical result of a time of flight experiment (circles), with a fit to  $\sigma_x^2(\tau)=\sigma_x^2(\tau=0)+\sigma_v^2\tau^2$ (solid line). (d) Size of the atomic cloud. Obtained by scanning the number of atoms, and hence density, in the trap and extrapolating to zero density using microwave pulses (blue circles, see text). Solid line is a linear fit. The data presented in (b-d) corresponds to measurements at thermal equilibrium, yielding the value of $\chi_H=0.82\pm0.02$.}
    \label{fig:fig1}
\end{figure}
In the experiment [Fig.~\ref{fig:fig1} (a)], a cloud of \Rb atoms is magneto-optically trapped and then cooled down to a temperature of $\sim20\mu$K. The final cooling step is optical evaporation in a far detuned, 1064~nm crossed dipole trap focused down to a waist of $60\mu$m overlapped with a strong, $\sim180$W, single 1070~nm beam tube trap (YLR-200-LP-AC-Y14, IPG photonics), loosely focused to a waist of $120\mu$m to provide strong confinement in the radial direction while leaving the axial dynamics practically unaffected. The Rayleigh length of the beam is $>4~$cm, much larger than any other relevant length scale. Extra care is taken to avoid reflections that may cause interference affecting the dynamics. The $>2$~sec long evaporation, much longer than the $\sim100$~msec collision time leaves the atoms in thermal equilibrium with the confining potential~\cite{Ketterle1996}. The atoms are then coupled for a duration $t$ to a non-thermal heat bath, implemented using a 1D dissipative Sisyphus lattice, where they may exhibit anomalous dynamics, depending on the modulation depth $U_0$~\cite{Sagi2012,Afek2017Correlations}. Set by the power and detuning of the lattice beams from the relevant atomic transition, $U_0$ is the main control parameter of the experiment. The other experimental parameters are similar to those in~\cite{Afek2017Correlations}. The trap depth is $\sim$~3~MHz, small compared to the $\sim60$~MHz detuning of the lattice, rendering trap-induced differential AC stark shifts negligible. Each measurement of the equipartition parameter $\chi_H(U_0,t)$ is comprised of three separate experiments: trap oscillations, time of flight and extrapolation to zero density of an in situ image of the cloud, giving access to the information needed to calculate the equipartition parameter of Eq.~\ref{eq:chi_definition}. The in situ absorption images are taken as the cloud is released from the trap, and capture both the atoms trapped in the focus, those held in the area of the beams removed from the overlapped focii. To ensure we do not wrongly include these atoms in our analysis, we fit the data to a sum of two Gaussians and use the narrower one~\footnote{See supplementary material}.\\

Fig.~\ref{fig:fig1} (b) shows a typical trap oscillation experiment, in which center of mass oscillations are excited using a short, near resonant light pulse. The atoms are sequentially imaged as a function of the time elapsed after the pulse. The measured frequency is $\omega=2\pi\times(332\pm2)$~Hz (used throughout the Paper) with approximately 1.6 oscillations before $1/e$ decay of the contrast, attributed to dephasing of the ensemble-averaged oscillations due to the anharmonicity of the confining potential. The trap frequency itself is unaffected by anharmonicity for atoms much colder than the trap depth.

The width of the velocity distribution is measured using time of flight. The cloud is released and allowed to expand in one dimension for a time $\tau$ within the tube trap. Its size is fitted with the relation $\sigma_x^2(\tau)=\sigma_x^2(\tau=0)+\sigma_v^2\tau^2$ between the standard deviation of the spatial distribution $\sigma_x(\tau)$ and that of the velocity distribution $\sigma_v$. Fig.~\ref{fig:fig1} (c) shows the result of such a measurement, giving $\sigma_v=42\pm1$ mm/sec. We verify that scattered light from the unidirectional tube trap does not affect the dynamics substantially by allowing the atoms to expand in it for a relatively long time ($>100$~msec, much longer than the duration of the experiment) and verifying that the center of mass of the cloud does not shift due to scattering of trap photons by more than 1\% compared to its initial position.

Measuring the in situ cloud size is challenging, mostly due to the optical density of the clouds, biasing the output of our absorption imaging. To alleviate this, we excite a controlled, variable fraction of the atoms homogeneously into the $F=2$ hyperfine state using a microwave pulse, scanning the density of the atoms in a given trap while leaving the density profile unchanged~\cite{Tung2010}. The transferred atoms are imaged using state selective absorption imaging and the cloud size extracted from fitting the distribution. The obtained values are then fitted with a linear relation and extrapolated down to zero atoms, representing zero density as the trap does not change. This gives the unbiased cloud size. Fig.~\ref{fig:fig1} (d) shows a result of such a measurement, yielding $\sigma_x(t=0)=24.5\pm0.3~\mu$m. Combining these we get, for atoms in thermal equilibrium after a long period of evaporative cooling, $\chi_H=0.82\pm0.02$, where the error is of statistical origins. Considering the possible systematics in such a measurement, one needs to look at the factors affecting each of the three measurements. Time of flight is a well-established technique, growing in precision as the cloud is allowed to expand more and more compared to its initial size. Our ``tube" trap allows these long measurement times as it prevents the atoms from expanding in the radial direction. We have also compared the time-of-flight results to a spectroscopic Raman velocity selective measurement, obtaining agreement to within 5-10\%. Trap oscillations are measured very precisely and with small errorbars. This can also be verified independently by looking at revival periods of quantum coherence imprinted with Raman control and is in very good agreement~\cite{Afek2017Revivals}. Initial cloud size is a challenging measurement due to systematics arising from high densities in the trap and imaging resolution. To further verify this we performed yet another measurement, scanning the density in a different way. Instead of using a MW pulse to homogeneously transfer atoms to the excited state and imaging them, we scan the initial MOT power and effectively begin the experiment with a variable number of atoms in the same sized dipole trap. The results agree well, to within less than 5\%. The fact that anharmonicity in a typical optical trap is much more substantial than simple Gaussian corrections has been previously established and verified with spectroscopic tools, much more accurate and susceptible to different types of systematics~\cite{Afek2017Revivals}. We can therefore conservatively bound the systematic error on our measurement of $\chi_H$ from above at the ~10\% limit. The deviation of this result from the theoretical unity value is related to the anharmonicity of the confining potential and will be elaborated later on.

\subsection*{Dynamics and steady-state}

\begin{figure} 
    \centering
        \begin{overpic}
            [width=\linewidth]{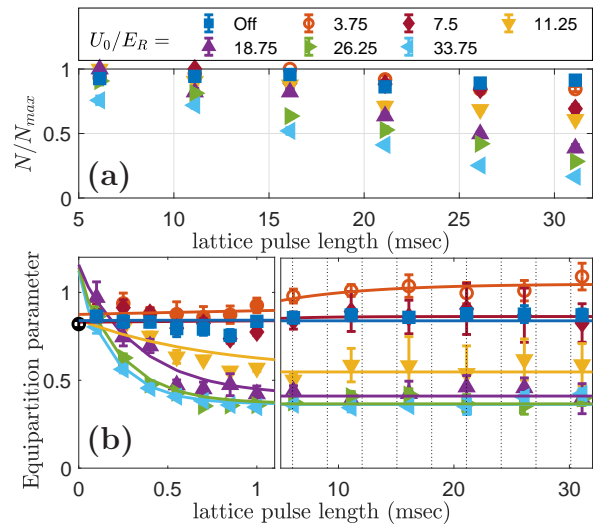}
            \put(35,130){\large \textbf{(a)}}
            \put(35,30){\large \textbf{(b)}}
        \end{overpic}
        \caption[Dynamics of equipartition.]{Dynamics of the equipartition parameter $\chi_H$. (a) Number of remaining atoms in the trap as a function of exposure time and lattice depth of the Sisyphus lattice. (b) The equipartition parameter per Sisyphus lattice depth and exposure time. Left (right) panel corresponds to short (long) times. Blue squares correspond to thermal dynamics, colored markers to anomalous dynamics. The black point at $t=0$ represents the value obtained in Fig.~\ref{fig:fig1} (b-d), according to Eq.~\ref{eq:chi_definition}. Solid lines are exponential fits (see text). Dotted vertical lines are integer multiples of the trap oscillation period.} 
    \label{fig:fig2}
\end{figure}

Fig.~\ref{fig:fig2} describes the dynamics of the equipartition parameter of the ensemble $\chi_H$ under coupling to both the confining potential and the non-thermal heat bath, compared to the case of thermal equilibrium with the optical trap. Fig.~\ref{fig:fig2} (a) shows the number of atoms remaining in the trap as a function of the lattice exposure time and depth. For deep lattices losses are substantial (up to a factor of about ten), mostly due to radial heating from photon scattering in the directions orthogonal to the lattice beams~\cite{Sagi2012}.  

\begin{figure*} 
    \centering
        \begin{overpic}
            [width=0.8\linewidth]{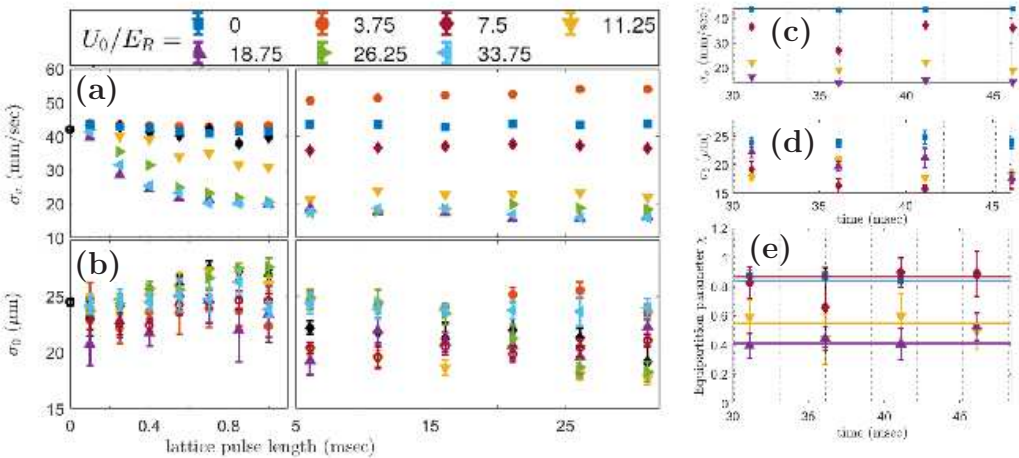}
            \put(37,150){\large \textbf{(a)}}
            \put(37,84){\large \textbf{(b)}}
            \put(300,172){\large \textbf{(c)}}
            \put(300,130){\large \textbf{(d)}}
            \put(292,90){\large \textbf{(e)}}
        \end{overpic}
        \caption[Dynamics of equipartition.]{RMS position (a) and velocity (b) as a function of exposure time and and lattice depth of the Sisyphus lattice for the data shown in Fig.~\ref{fig:fig2}. Panels (c-e) are the long-time measurements of position, velocity and $\chi_H$ taken for a select number of lattice-depths, where atom loss was not a limiting factor in the measurement.}
    \label{fig:fig3}
\end{figure*}

The dynamics of the equipartition parameter is given in Fig.~\ref{fig:fig2} (b), taking into account the trap oscillation frequency measured in Fig.~\ref{fig:fig1} (b) and calculating $\chi_H(U_0,t)$ according to Eq.~\ref{eq:chi_definition}. Each point is comprised of a set of the three experiments described above, sampled 20 times at a random order. Error bars are evaluated by considering the 67\% confidence intervals of the linear fits used to determine $\sigma_x(t=0)$ and $\sigma_v$. Solid lines are fits performed by taking the short time data $t<1.2~$msec and fitting it to $A\exp(-\gamma t)+C$ to get the decay rate $\gamma$. Then, all the data is included and fitted to $(\chi_0-\chi_\infty)\exp(-\gamma t)+\chi_\infty$, with the $\gamma$ from the short time fit. The fact that this is indeed a steady state value and not a transient effect is proven by waiting many trap oscillation periods, shown in dashed vertical lines in Fig.~\ref{fig:fig2} (b, right panel). Density dependent effects such as $s$-wave atomic collisions and light assisted repulsion~\cite{Wieman1989} can be ruled out since despite the fact that the number of remaining atoms in the trap, and hence the density of the confined atoms, changes by up to an order of magnitude, smooth behavior of the equipartition parameter is observed. The fast timescale of the dynamics of $\chi_H$ is determined primarily by that of the velocity dynamics. Our simulations indicate that the slower timescale is related to that of the relaxation of the position distribution. Note that for shallow lattices the dynamics becomes extremely slow~\cite{Hirschberg2011,Hirschberg2012}, hence for the $U_0/E_R=3.75$ dataset (red circles) there is no obvious steady state achieved within the duration of the experiment. We use $E_R$, the recoil energy for the \Rb $D_2$ line as the relevant energy scale. Figure~\ref{fig:fig3} shows the RMS position and velocity [(a) and (b) respectively] for the data shown in Fig.~\ref{fig:fig2}. Except for the deepest lattices, the relevant time scale for reaching the non-equilibrium steady state is that of the trap oscillation period. We are able to evolve the system for $\sim10$ oscillation periods allowing all thermodynamic variables to reach steady state. This is even further verified in panels (c-e) of Fig.~\ref{fig:fig3}, showing additional $\sim5$ oscillation periods for mid-range lattice depths.

\begin{figure} 
    \centering
        \begin{overpic}
            [width=\linewidth]{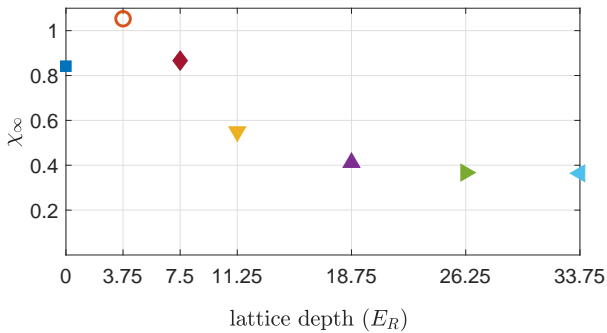}
        \end{overpic}
        \caption[Equipartition.steady state]{Steady state equipartition parameter as a function of lattice depth [extracted from the exponential fit of Fig.~\ref{fig:fig2} (b)]. Error bars are smaller than marker size. High Sisyphus lattice powers significantly decrease the values of $\chi_H$ from their thermal value.}
    \label{fig:fig4}
\end{figure}

We summarize the steady state values of the harmonic equipartition parameter as a function of lattice depth in Fig.~\ref{fig:fig4}. The thermal value corresponds to the lattice being turned off. As the lattice gets deeper, the steady state value is reduced, in accordance with the theoretical prediction~\cite{Dechant2015,Dechant2016}. The value of the equipartition parameter was predicted to begin to climb back towards its thermal value with further increase of the depth of the lattice, however we were not able to observe this behaviour due to experimental constraints. The steady state $\chi_H$ values for shallow lattices are higher than the equilibrium value. We attribute these deviations to residual heating of the atoms at shallow Sisyphus lattices~\cite{Castin1991,Katori1997,Douglas2006,Sagi2012}, in consistency with Eq.~\ref{eq:chi_definition}.

\subsection*{Effects of anharmonicity of the confining potential}

\begin{figure} 
    \centering
        \begin{overpic}
            [width=\linewidth]{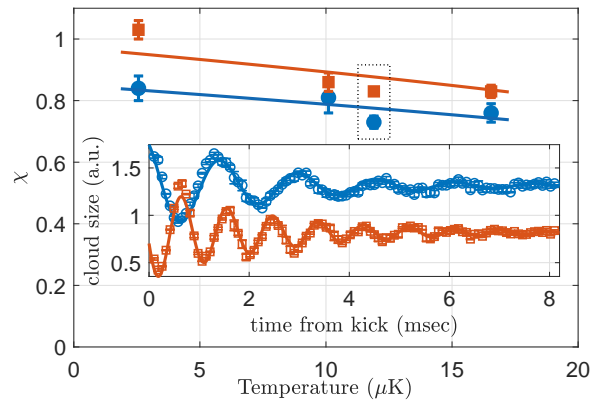}
        \end{overpic}
        \caption[Effect of anharmonicity]{(a) $\chi_H$ as a function of the temperature, scanned by varying the depth of the final optical evaporative cooling stage, of a 3D trapped ensemble for two axes: horizontal (blue circles) and vertical (red squares). The inset shows a measurement of the trapping oscillations, taken for the $\chi_H$ values in the dotted rectangle. More oscillations prior to substantial decay are observed in the vertical axis, indicating higher harmonicity. Solid lines are best fits to the proportionality relation for $\chi_H$ in a Gaussian trap described in the text.}
    \label{fig:fig5}
\end{figure}

We now return to the interpretation of the deviation of the measured thermal value of $\chi_H=0.82\pm0.02$ from unity (Figs.~\ref{fig:fig1},~\ref{fig:fig2}). Deviations of $\chi_H$ from unity are caused by two independent factors. The first is the anharmonicity of the confining potential. It has been recently shown~\cite{Afek2017Revivals} that the anharmonicity of the confining potential plays a pivotal role in determining the dynamical properties of the system. To see how it affects the equipartition parameter, we perform a measurement of $\chi_H$ as a function of temperature, obtained by varying the depth of the final optical evaporative cooling stage. We do so for both axes of the trap, horizontal (in which the main experiment is performed) and vertical. The horizontal axis suffers from greater anharmonicity due to residual trapping of atoms outside of the crossed region of the dipole trap (``wings"). This is manifested in a substantial decay that occurs after a smaller number of trap oscillations compared to that of the vertical axis. The effect, depicted in Fig.~\ref{fig:fig5}, is twofold: As the temperature is lowered, the ratio of the energy of the atoms and the depth of the trap is reduced and the atoms sample less anharmonicity, bringing about an approach of the equipartition parameter to unity for both axes. The vertical axis gives a higher equipartition parameter throughout the temperature range, attributed to the fact that the anharmonicity there is inherently lower. In the experiment we measure $\chi_H$, rather than $\chi_G$, a Gaussian equipartition parameter that can be obtained directly from Eq.~\ref{eq:equipartition}, due to experimental considerations. Assuming a purely Gaussian type of anharmonicity, described by the Hamiltonian $\Ha=p^{2}/(2m)-A\exp\left[-x^{2}/(2\sigma_{T}^{2})\right]$, where $A$ in the depth of the potential and $\sigma_T$ its width, and defining $\alpha\equiv\sigma_x/\sigma_T$, the size of the atomic distribution relative to the size of the trap and $\beta\equiv k_BT/A=\alpha/(1+\alpha)^{3/2}$, the temperature of the atoms relative to the depth of the trap, one can derive analytically~\cite{Note1} a scaling relation between $\chi_H$ and the temperature, $\chi_H=\sqrt{\beta/\alpha}$. We use this relation to fit the data of Fig.~\ref{fig:fig5}, to good agreement up to a separate initial value for $\chi_H(T=0)$ used as a fit parameter to account for different inherent anharmonicity between the axes.

\begin{figure} 
    \centering
        \begin{overpic}
            [width=\linewidth]{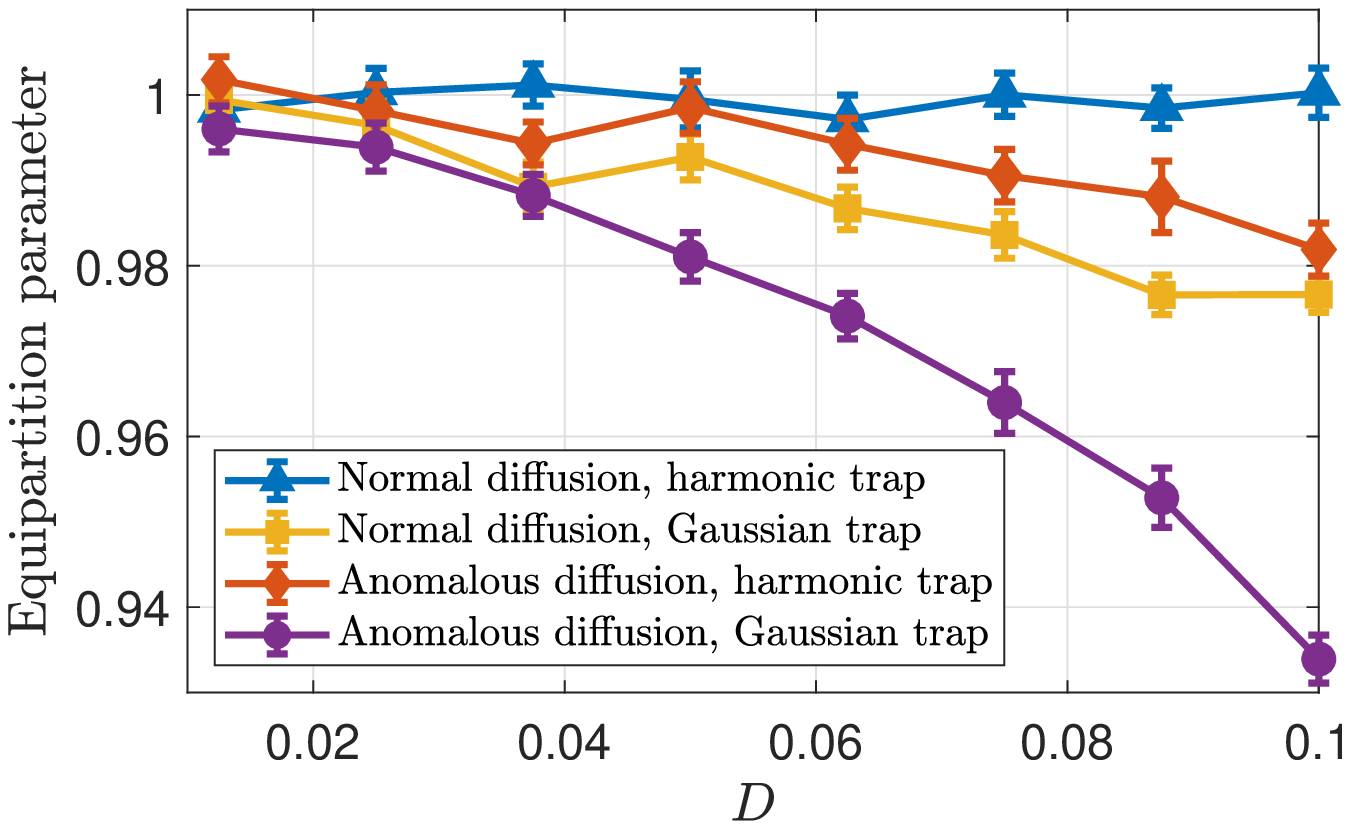}
        \end{overpic}
        \caption[Effect of anharmonicity]{Monte-Carlo simulation of steady state $\chi_H$ for harmonic and Gaussian anharmonic trapping potentials, as a function of the dimensionless diffusion coefficient $D$. For normal diffusion (thermal equilibrium, no Sisyphus lattice, $D\sim T$, the temperature of the atoms) in harmonic trap (blue triangles), equipartition holds for all values of the diffusion coefficient. In an anharmonic trap, the hotter the atoms are, the more anharmonicity they experience, increasing the deviation from unity. Anomalous diffusion (non-thermal equilibrium, Sisyphus lattice on, $D\sim E_R/U_0$, the inverse lattice depth) generates deviations from unity even for harmonic potentials (red diamonds). The effect of anomalous dynamics combined with anharmonic potential is additive in the decrease of $\chi_H$ (purple circles).}
    \label{fig:fig6}
\end{figure}

Fig.~\ref{fig:fig6} presents numerical simulations comparing between values of $\chi_H$ for normal diffusion, \ie thermal equilibrium and no Sisyphus lattice, for harmonic (blue triangles) and anharmonic Gaussian (gold squares) traps. The horizontal axis is the dimensionless diffusion constant that for normal diffusion is proportional to the temperature of the atoms. For a harmonic trap $\chi_H=1$ for all temperatures. For the anharmonic trap, as the temperature increases the atoms sample more anharmonicity and the harmonic equipartition parameter diminishes. This is the effect we associate with our thermal equilibrium result. The reason the effect in the experiment is more pronounced is that Gaussian anharmonicity does not suffice to describe the real anharmonicity typical for dipole traps~\cite{Afek2017Revivals}. It is still, however, very useful for simplification of calculations and qualitative analysis. The second factor contributing to the decay is the main result of this Paper. The predicted behavior for harmonic potential~\cite{Dechant2015,Dechant2016} is reobtained in our simulations, using the reduced semi-classical Sisyphus cooling mechanism in the regime of deep lattices, where the dimensionless diffusion coefficient is $\sim E_R/U_0$ (Fig.~\ref{fig:fig6}, red diamonds). Finally, we show (purple circles) that the two effects are additive, confirming our experimental results and showing that the breakdown of equipartition persists in anharmonic potentials and $\chi_H$ is a fair predictor for it~\footnote{Within the semi-classical model~\cite{Castin1991,Marksteiner1996}, one can come up with a unit transformation under which $D \sim E_R/U_0$, the dimensionless diffusion constant, is proportional to the inverse depth of the lattice (full derivation and interpretation of other parameters given in~\cite{Note1}). For normal diffusion, under the same unit transformation $D$ is the usual momentum diffusion coefficient given by the Einstein relation $D_{p}=\gamma mk_{B}T$, proportional to the temperature of the bath}. Our simulations further indicate a violation of equipartition for yet another interesting class of anharmonic perturbations - logarithmically corrected harmonic potentials~\cite{Note1}. Agreement between the experiment and theoretical predictions is qualitative. This is not new for this system~\cite{Sagi2012,Afek2017Correlations} and is mostly attributed to the complex atomic level structure ignored by the semi-classical model of Sisyphus cooling.

\subsection*{Inhomogeneous distributions of kinetic energy}

\begin{figure} 
    \centering
        \begin{overpic}
            [width=\linewidth]{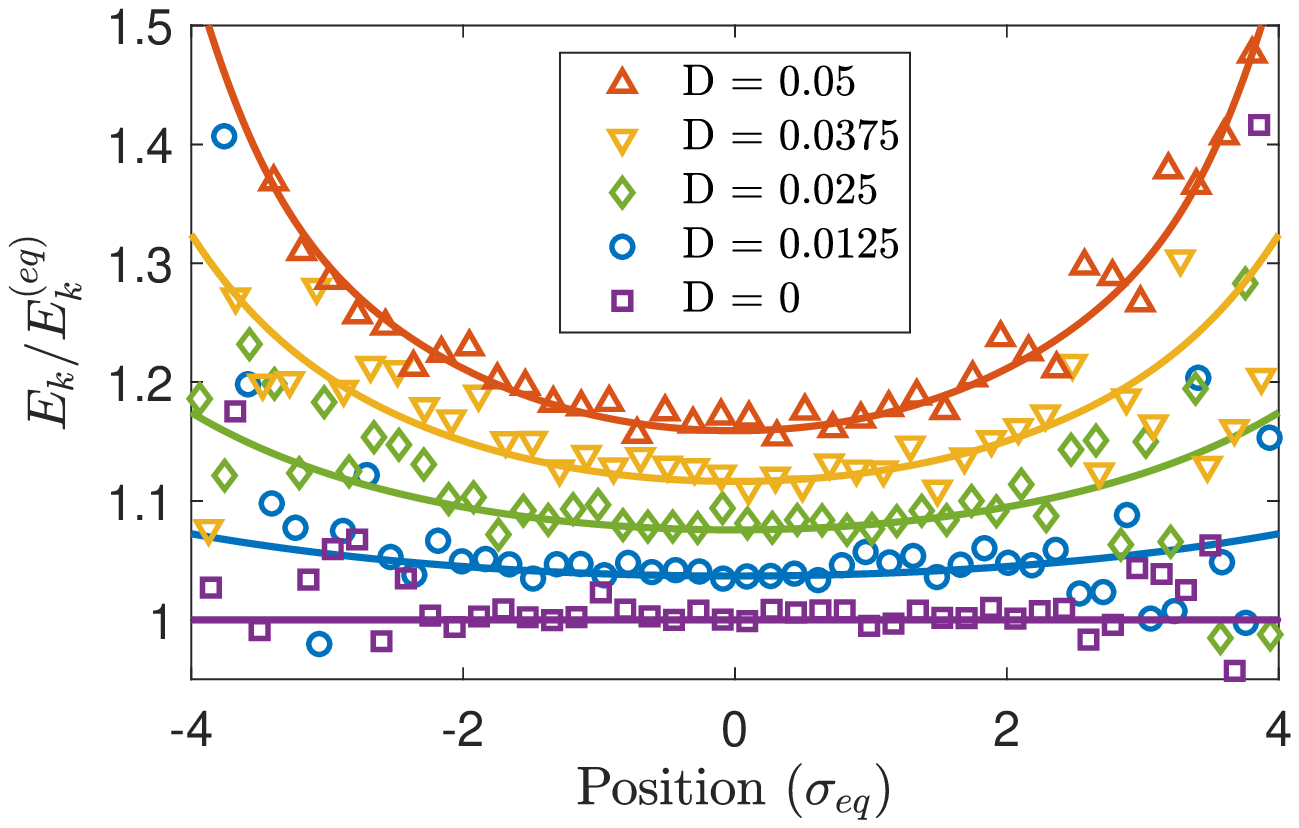}
        \end{overpic}
        \caption[Gradients of kinetic and potential energies]{Dependence of the kinetic energy, normalized by its equilibrium value, on position, normalized by equilibrium Gaussian RMS value, at steady state of anomalous dynamics in a harmonic trap, calculated independently numerically (markers) and analytically (solid lines). The case of $D=0$ represents thermal equilibrium normal diffusion. For higher $D$ values the local kinetic energy shows a stronger dependence on position~\cite{Note1}.}
    \label{fig:fig7}
\end{figure}

The steady-state phase space representation of the system has been theoretically studied in~\cite{Dechant2016}. It was found that equivalence between equi-energetic and equi-probable surfaces no longer holds. Another fascinating aspect can be revealed by studying correlations of the kinetic energy, $\sim v^2$, with position. Specifically, the kinetic energy is found to be inhomogeneous, \ie position dependant. In Fig~\ref{fig:fig7} we present the results of our analytics and numerics, using the methods described in~\cite{Dechant2016}, The local average kinetic energy was calculated as the marginal expectation value of the kinetic energy term with respect to the total phase space probability distribution function at a fixed position. Both methods are in good agreement. Notice that different units are in Figs.~\ref{fig:fig6} and~\ref{fig:fig7}, resulting in a slightly different interpretation of $D$~\cite{Note1}. The results imply inseparability of the phase space probability distribution function, in contrast to thermal distributions, and enhanced kinetic-potential energy correlations. The inhomogeneity of the kinetic energy at a steady state can serve as direct evidence of the non-thermal nature of the Sisyphus dissipative lattice. 

In recent work~\cite{Afek2017Correlations}, we put forth a technique enabling direct imaging of the phase-space density distribution function of an atomic ensemble. Utilizing a higher-order version of such a method, studying correlations of the kinetic energy, $\sim v^2$, with position, it may be possible to observe this position dependence of the kinetic energy, testing our prediction.

\subsection*{Summary and outlook}

In summary, we presented a detailed experimental observation of the previously overlooked deviation from generalized equipartition in dilute, confined, laser cooled atoms, looking not only at steady-state behaviour but also at the dynamics. We introduced the equipartition parameter, which can serve for quantifying the departure from thermal equilibrium of non-thermal states and established its relation to the anharmonicity of the confining potential. With improved signal to noise it should be interesting to attempt a direct measurement of $\chi$ (Eq~\ref{eq:chi}) involving the full details of the confining potential. Finally, we presented a new prediction of inhomogeneous kinetic and potential energies for the system of confined, laser cooled atoms, supported by analytical and numerical methods and experimentally attainable.

\begin{acknowledgments}
The authors would like to thank Eli Barkai, Andreas Dechant, David Mukamel and Oren Raz for fruitful discussions.
\end{acknowledgments}

\bibliographystyle{apsrev4-1}
\bibliography{equipartition}

\end{document}



\heading{Supplementary Material for Deviations from generalized Equipartition in Confined, Laser Cooled Atoms}
\begin{center} Gadi Afek\footnote{Present address: Wright Laboratory, Department of Physics, Yale University, New Haven, Connecticut 06520, USA}, Alexander Cheplev, Arnaud Courvoisier and Nir Davidson\end{center}
\begin{center} Department of Physics of Complex Systems, Weizmann Institute of Science, Rehovot 76100, Israel\end{center}


\section{Experimental}

\subsection*{Details of the fitting procedure}

\begin{figure}[H]
    \begin{centering}
    \includegraphics[width=\textwidth]{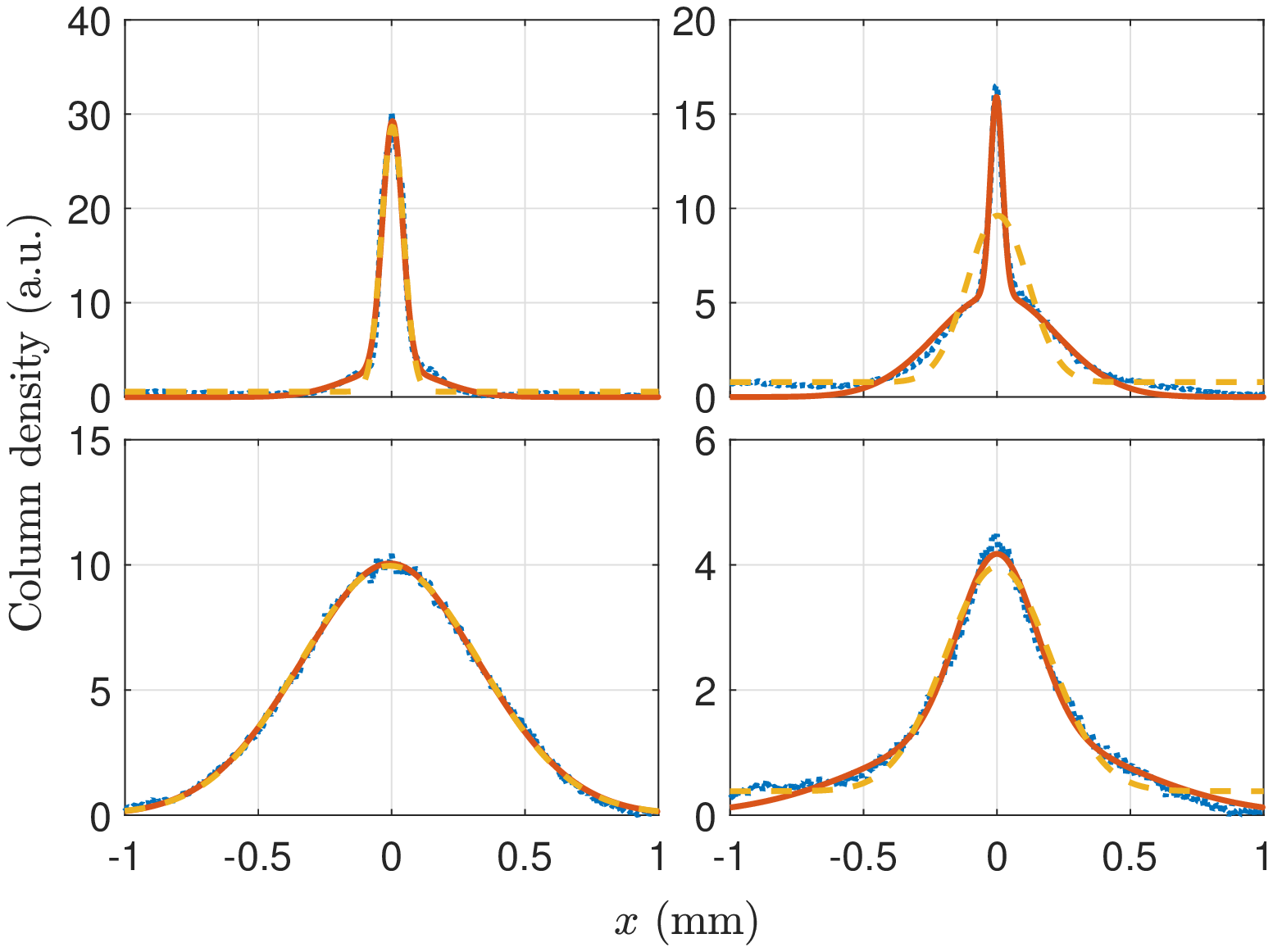}
    \par\end{centering}
    \caption{The integrated density profiles of the horizontal axis obtained from the images used in the main text. The left column represents data for $U_0/E_R=0$, meaning that the lattice is turned off. The right column represents the data for $U_0/E_R=11.25$. All figures are for lattice exposure time of 31.1~msec. The top row is taken from in-situ images and the bottom row is taken from 7.1~msec time of flight beginning after the 31.1~msec lattice exposure. In all the images the data (blue dotted line) is compared with the double Gaussian fit (red solid line) and a naive Gaussian fit (orange dashed)\label{fig:Fit}}
\end{figure}

In the main text we describe results of fitting the absorption images of the cloud to a double (bimodal) Gaussian. This is a necessary procedure since the measurement requires in situ imaging, highly affected by the atoms trapped in the ``wings" of the trap. In order to be able to extract the standard deviation of only the ``interesting" atoms (albeit compromising the ability to probe the tails of the distributions), trapped within the crossed part of the dipole trap, we fit the data in the following way: We attempt to fit a sum of two Gaussians. If the output is two Gaussians with $|\sigma_1/\sigma_2-1|<0.08$ (meaning that a single Gaussian would do just as well) we fit with a single Gaussian. Then for all further analyses we use the thinner Gaussian, representing the atoms trapped in the focus and not in the wings. Figure~\ref{fig:Fit} shows the integrated density profile of the horizontal axis for two scenarios: one where the lattice is turned off ($U_0/E_R=0$, or simply doing nothing) in the left column and another where the lattice depth is set to $U_0/E_R=11.25$ shown in the right column. In both cases the elapsed time from the beginning of the experiment (end of evaporation) is 31.1~msec. The bottom panels show the distributions after a 7.1~msec time of flight initiated after the 31.1~msec lattice exposure has elapsed. The data is fitted with both a double Gaussian and a single (naive) Gaussian.

Even for $U_0/E_R=0$, in the in situ image (top left panel) there are considerable deviations from a single Gaussian, biasing the naive fit. For obtaining the second moment of the distribution of the atoms trapped only in the focus it suffices to use the thin Gaussian obtained by the bimodal fit. The wings are no longer visible after a time of flight (bottom left panel), rendering the data effectively indistinguishable from both fits. The situation is aggravated for the anomalous dynamics of the lattice plus confinement system. In the in situ profile (top right panel) the naive fit clearly fails, but the bimodal catches the essence of the width of the important atoms. Note that the velocity distribution for this case is inherently non-Gaussian~\cite{Katori1997,Jersblad2004,Douglas2006}. However, for low moments, far from the tails of the distribution, the Gaussian fit is still useful. In the example of the bottom right panel of Fig.~\ref{fig:Fit}, both fits give moments that are similar to the moments calculated directly from the data to within 15\% for moments of order lower than $\sim2.5$. The simulations also support the use of Gaussian fits for the distributions at hand as far as low moments are concerned (see Fig.~\ref{fig:Gaussian-fits-example} and relevant discussion).

\section{Numerical Methods and Results}

The simulation results presented both in the main text and this supplementary material were obtained by Monte-Carlo simulation of appropriate Langevin equations, by method of Euler\textendash Maruyama. The Langevin equations that correspond to semi-classical description of anomalous diffusion of cold atoms in a dissipative optical lattice, confined by a harmonic potential are given by \cite{castin1991limits} -
\begin{align*}
\dot{x} & =\f pm\\
\dot{p} & =-\f{\gamma p}{1+\left(p/p_{c}\right)^{2}}-m\omega^{2}x+\sqrt{2D_{p}}\xi\left(t\right)\\
D_{p} & =D_{0}+\f{D_{1}}{1+\left(p/p_{c}\right)^{2}}
\end{align*}
Here $\xi\left(t\right)$ is Gaussian white noise, satisfying $\ex{\xi\left(t\right)\xi\left(t'\right)}=\delta\left(t-t'\right)$. The so-called reduced semi-classical model is obtained by neglecting the momentum dependent diffusion coefficient, effectively setting $D_{1}\to0$. We then follow~\cite{Dechant2015,Dechant2016} and change the variables to dimensionless units by setting $x=m\omega\t x/p_{c}$ and $p=\t p/p_{c}$ (tilde marks dimentionful quantities), and defining the dimensionless diffusion coefficient $D=D_{0}/\gamma p_{c}^{2}$ and dimensionless trap frequency at the bottom of the confining potential $\Omega=\omega/\gamma$. The resultant Langevin equations are -
\begin{align*}
\dot{x} & =\Omega p\\
\dot{p} & =-\f p{1+p^{2}}-\Omega x+\sqrt{2D}\xi
\end{align*}
For comparison, The Langevin equation for normal diffusion in these units are -
\begin{align*}
\dot{x} & =\Omega p\\
\dot{p} & =-p-\Omega x+\sqrt{2D}\xi
\end{align*}
The difference between the models thus is by different friction mechanisms fed by fluctuations drawn from identical Gaussian distribution. This facilitates the comparison between the models through the parameter $D$, which for the thermal case is proportional to the bath temperature through the Einstein relation, $D_{p}=\gamma mk_{B}T$ and for the anomalous diffusion effectively quantifies the magnitude of the anomalous dynamics. For the case of Gaussian confining potential, the force term due to confinement was replaced from $-\Omega x$ to $-\Omega x\exp\left(-\f{x^{2}}{2\sigma_{t}^{2}}\right)$, where $\sigma_t$ is the RMS size of the trapping potential. The results presented in Fig.~7 of the main text were obtained after a second unit transformation of the Langevin equations, by rescaling the variables $z=x/\sqrt{D}$ and $u=p/\sqrt{D}$ resulting in the following equations -
\begin{align*}
\dot{z} & =\Omega u\\
\dot{u} & =-\f u{1+Du^{2}}-\Omega z+\sqrt{2}\xi
\end{align*}
Notice that the definition of $D$ did not change, however for $D=0$ the equations reduce to these of normal diffusion, and for $D>0$ they manifest anomalous diffusion.

\subsection*{Effects of anharmonic trapping potentials}

\subsubsection*{Gaussian trap}

\begin{figure}[H]
\begin{centering}
\includegraphics[width=0.8\textwidth]{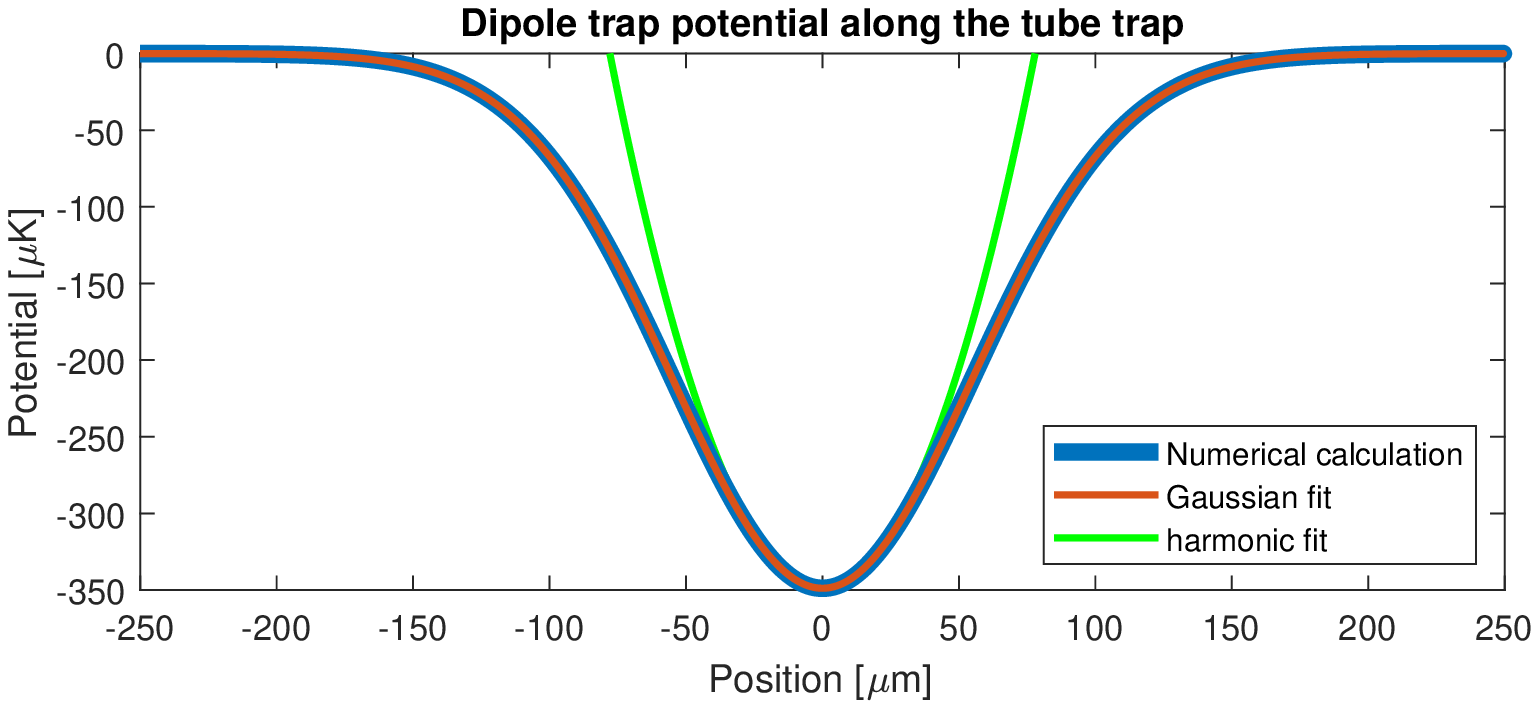}
\par\end{centering}
\caption{Numerical calculation of dipole trap potential along the tube trap, with Gaussian and harmonic fits. \label{fig:Dipole-trap-potential}}
\end{figure}

\begin{figure}[H]
\begin{centering}
\includegraphics[width=1\textwidth]{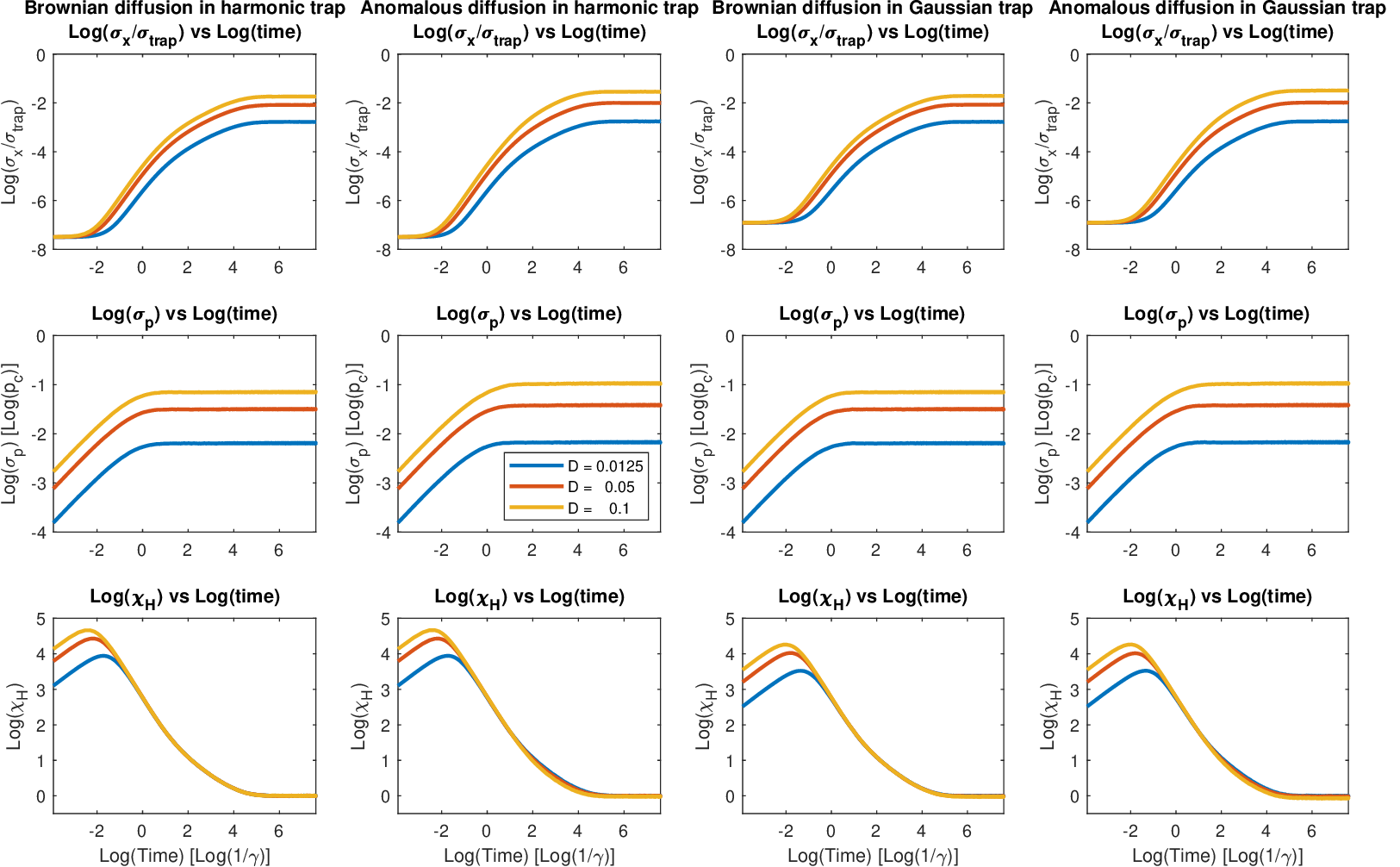}
\par\end{centering}
\caption{Time dependence of position and momentum RMS values for several settings and $D$ parameters, with the appropriate $\chi_H$ parameter that is calculated from them. \label{fig:Reaching-metastable-state}}
\end{figure}

\begin{figure}[H]
\begin{centering}
\includegraphics[width=0.6\textwidth]{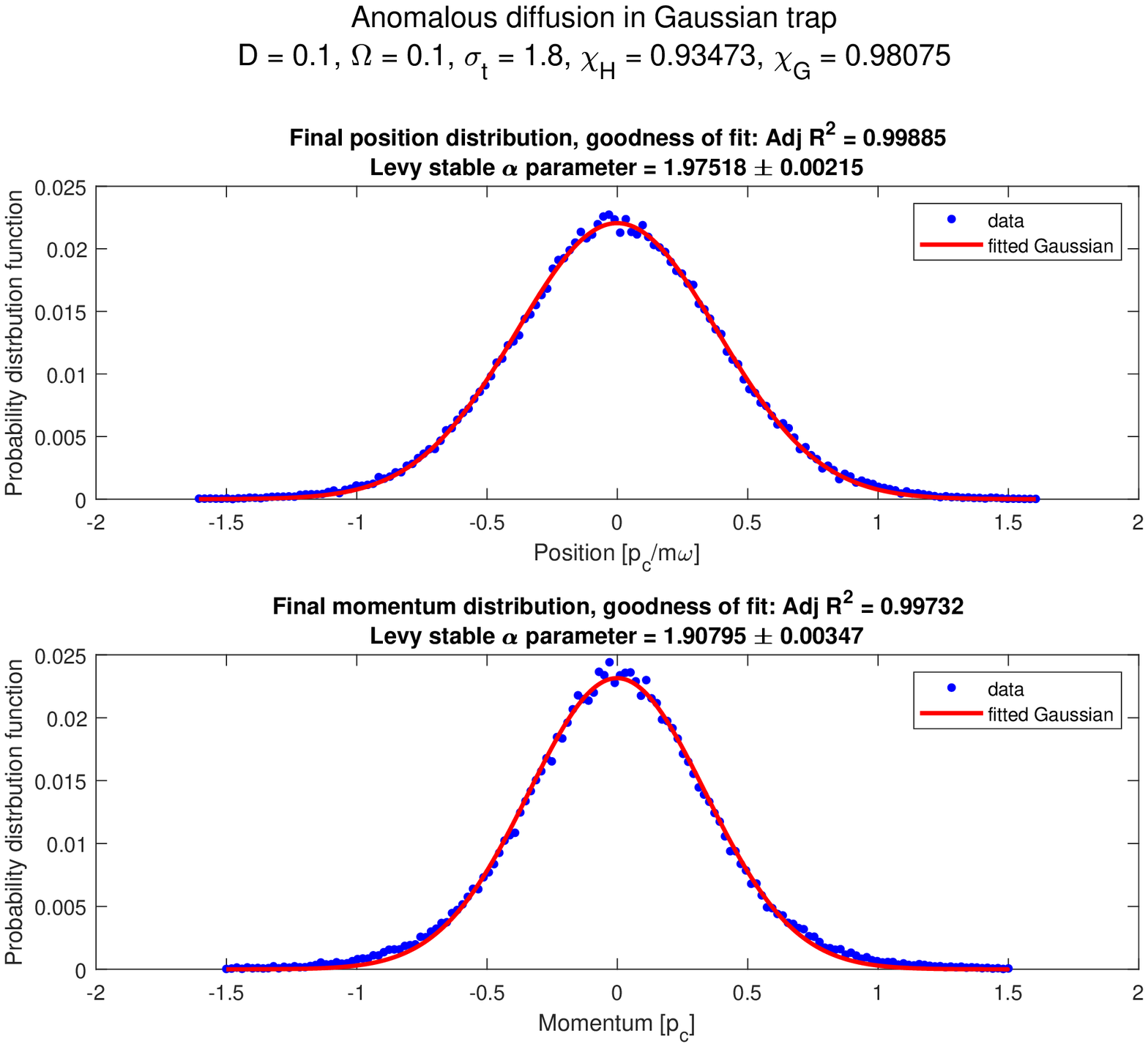}
\par\end{centering}
\caption{Example for Gaussian fits of position and momentum marginal distribution functions. \label{fig:Gaussian-fits-example}}
\end{figure}

\begin{figure}[H]
\begin{centering}
\includegraphics[width=0.8\textwidth]{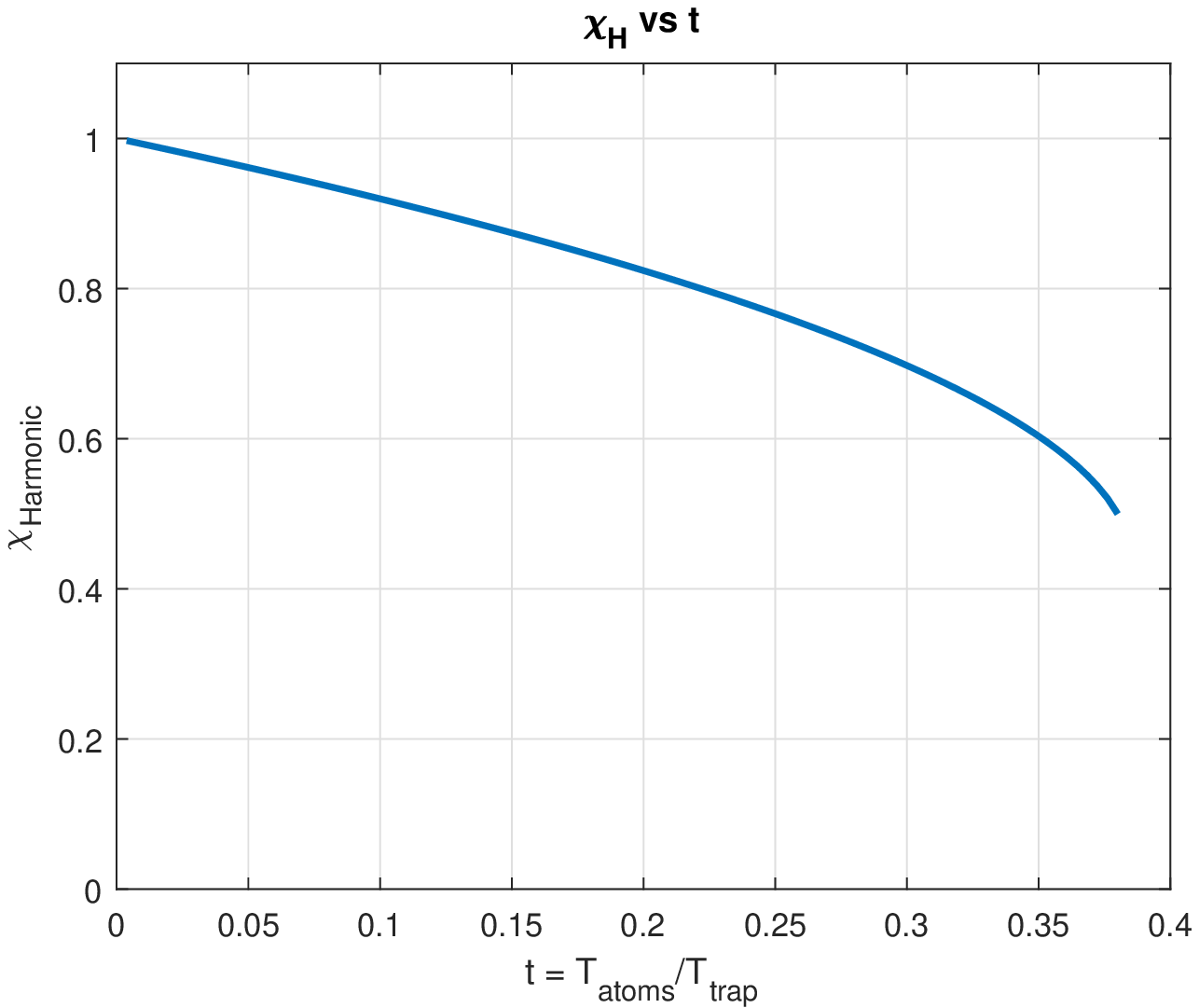}
\par\end{centering}
\caption{Analytical calculation of $\chi_{H}$ at Gaussian trap in thermal
equilibrium, under Gaussian distribution approximation. \label{fig:Analytical-calculation-of}}
\end{figure}

\begin{figure}[H]
\begin{centering}
\includegraphics[width=0.8\textwidth]{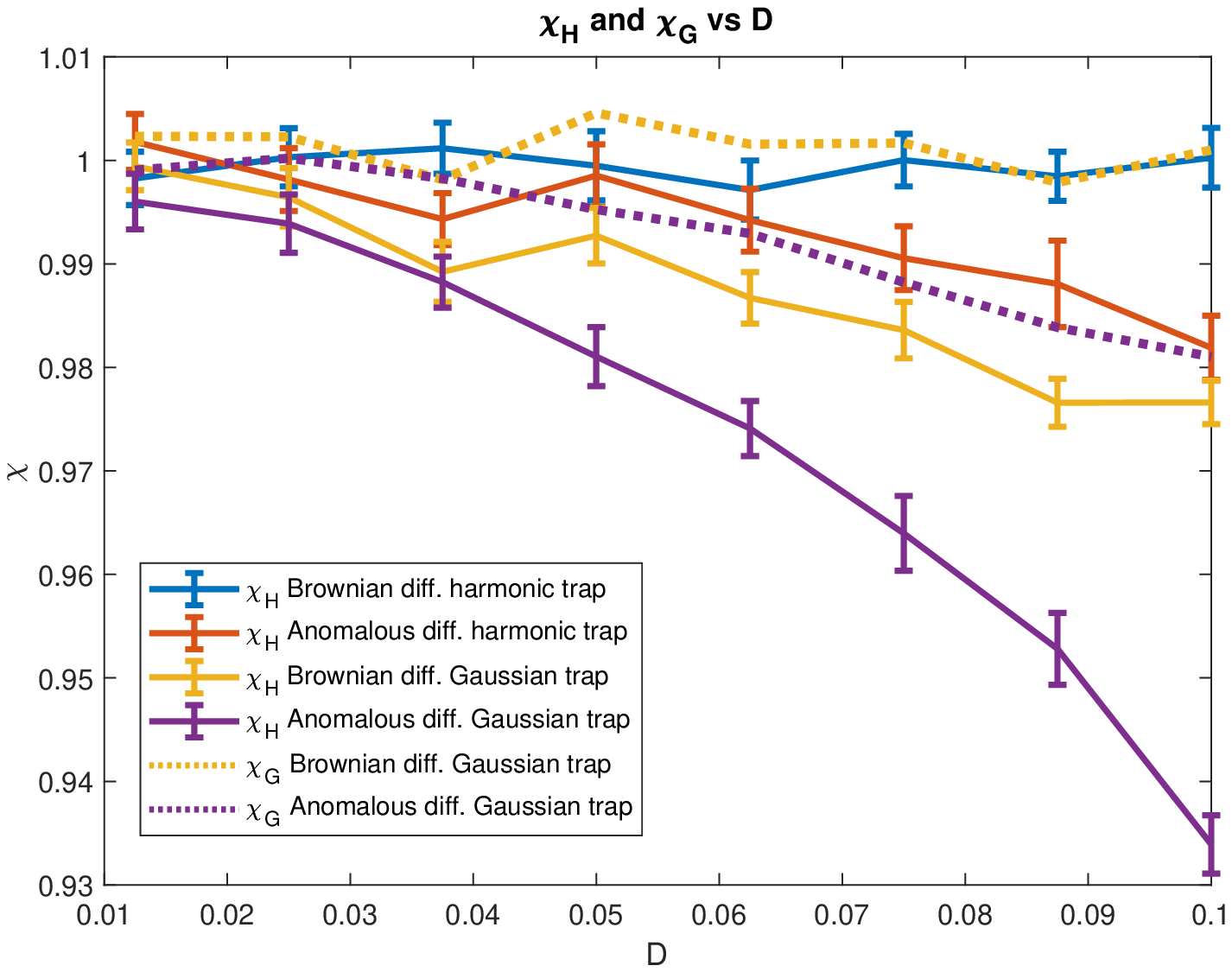}
\par\end{centering}
\caption{$\chi$ vs $D$ for harmonic and Gaussian potentials, for both Brownian and anomalous diffusion. \label{fig:chi_vs_D}}
\end{figure}

\begin{figure}[H]
\begin{centering}
\includegraphics[width=0.8\textwidth]{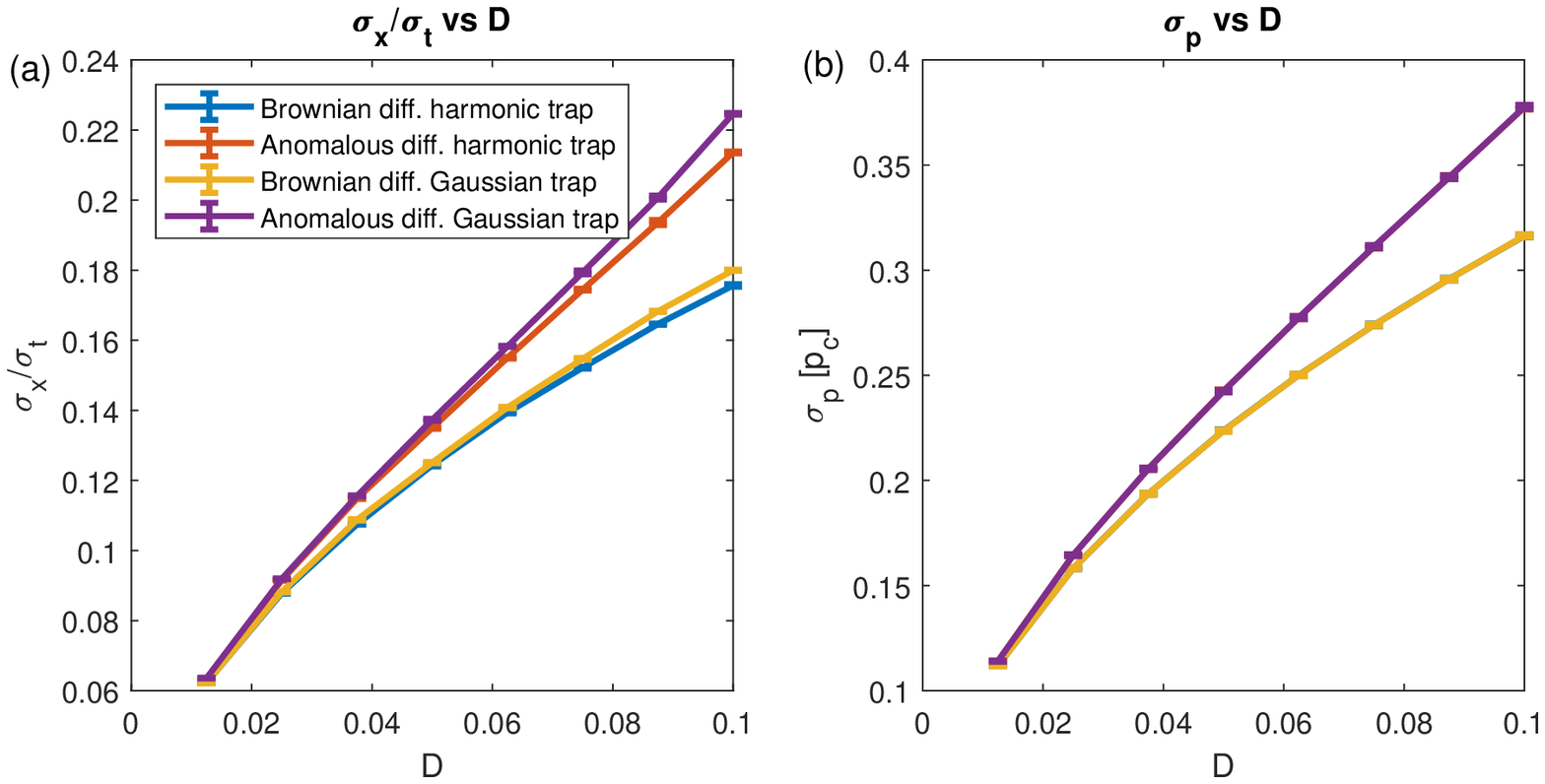}
\par\end{centering}
\caption{Position and momentum RMS values vs $D$ \label{fig:position-and-momentum_RMS_vs_D}}
\end{figure}

\begin{figure}[H]
\begin{centering}
\includegraphics[width=0.8\textwidth]{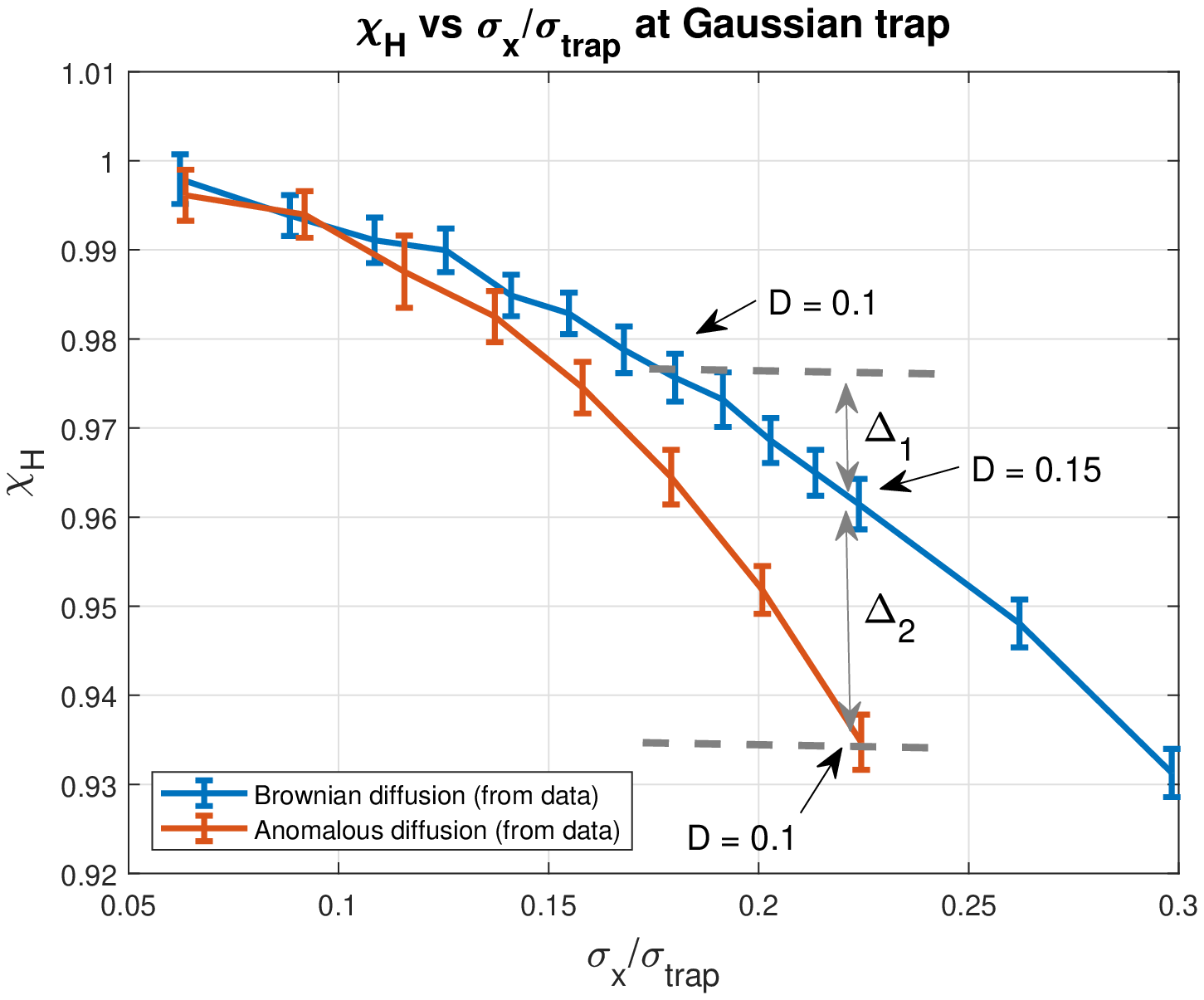}
\par\end{centering}
\caption{$\chi$ vs $\sigma_{x}$ as calculated directly from simulation data (without Gaussian fits) \label{fig:chi_vs_sigma_x}}
\end{figure}

As a prototype for confining anharmonic potential we used a 1D (inverted) Gaussian potential, $V\left(x\right)\sim-Ae^{-\f{x^{2}}{2\sigma_{t}^{2}}}$. This prototype model was used since the experimental confining potential is supported by Gaussian beams for the dipole trap. Using experimental geometry and beam parameters we calculate numerically the confining potential at along the axis of the experiment. The result, along with Gaussian fit and harmonic fit (of the bottom part of the potential) is presented in Fig.~\ref{fig:Dipole-trap-potential}, showing that a Gaussian model describes well our expected confining potential
Due to reasons such as optical aberrations, small variations of the experimental parameters and the overall geometry - the actual confining potential is expected to be similar to but not exactly Gaussian. Since the Gaussian potential in inherently meta-stable, it was crucial to verify that the Langevin simulations reached a long-lived meta-stable state. We considered ensembles with temperatures much smaller than the trap depth $A$, thus by Kramer's law\cite{Frank2005} enhancing exponentially the time that the atoms are confined within the trap. Fig.~\ref{fig:Reaching-metastable-state} shows how the position and momentum RMS values of the ensemble converge to their meta-stable values in Gaussian trap for Brownian and anomalous diffusion, and compares these to the case of harmonic confinement.
As was noted in the main text, the experimental data was fitted to a Gaussian function in order to suppress the effect of atoms trapped in the ``wings'' of the 3D physical confining potential. We test the validity of this fitting procedure using the simulation, and obtain a very good agreement in the simulated parameters regime. An example of such a fit is presented in Fig.~\ref{fig:Gaussian-fits-example}. 
Here $\chi_{H}$ is the harmonic equipartiton parameter, and $\chi_{G}$ is the ``correct'' equipartiton parameter for the Gaussian confining potential, derived from the Hamiltonian $\H=\f{p^{2}}{2m}-Ae^{-\f{x^{2}}{2\sigma_{t}^{2}}}$ (denoting $\omega_{G}=\sqrt{A/m\sigma_{t}^{2}}$ as the harmonic frequency at the bottom of the potential) -
\[
\chi_{G}=\sqrt{\f{\ex{p\pd{\H}p}}{\ex{x\pd{\H}x}}}=\f 1{m\omega_{G}}\sqrt{\f{\ex{p^{2}}}{\ex{x^{2}\exp\left(-\f{x^{2}}{2\sigma_{t}^{2}}\right)}}}
\]
The position and momentum distributions fit well to Gaussians in this regime of deep lattices (small $D$ values), however we still receive a considerable deviations form unity of the equipartiton parameter. Goodness of Gaussian fits can be exploited to predict the temperature dependence of the harmonic equipartiton parameter in Gaussian potential \emph{in equilibrium with a thermal bath}. Constraining the system to obey the Generalized equipartiton theorem, with $T_{a}$ being the temperature of the atoms - 
\begin{align*}
\f 1m\ex{p^{2}} & =k_{B}T_{atoms}\\
\f A{\sigma_{t}^{2}}\ex{x^{2}e^{-\f{x^{2}}{2\sigma_{t}^{2}}}} & =k_{B}T_{atoms}\\
\f{\f 1m\ex{p^{2}}}{\f A{\sigma_{t}^{2}}\ex{x^{2}e^{-\f{x^{2}}{2\sigma_{t}^{2}}}}} & =1
\end{align*} 
The expectation values are in regard to phase space probability distribution function. If we assume that the spatial distribution is well approximated by a Gaussian we get the following relation - 
\begin{align*}
\ex{\f 1{\sigma_{t}^{2}}x^{2}e^{-\f{x^{2}}{2\sigma_{t}^{2}}}} & =\int_{-\inf}^{\inf}\f 1{\sigma_{t}^{2}}x^{2}e^{-\f{x^{2}}{2\sigma_{t}^{2}}}\ubr{\f 1{\sqrt{2\pi}\sigma_{a}}e^{-\f{x^{2}}{2\sigma_{a}^{2}}}}{\text{probability dist.}\text{\ensuremath{\P\left(x\right)}}}\dif x\\
 & =\f{\alpha}{\left(1+\alpha\right)^{\f 32}}\\
 & =t
\end{align*}
Where we defined $\sigma_{a}$ as the width of the Gaussian spatial distribution of the atoms with $\alpha\equiv\left(\f{\sigma_{a}}{\sigma_{t}}\right)^{2}$ and $t\equiv\f{k_{B}T_{atoms}}A$ is the relative temperature of atoms with respect to the trap depth. We received the dimensionless polynomial version of the function $T_{a}\left(\ex{x^{2}}\right)$ - that is $t\left(\alpha\right)$ - what is the temperature given the position variance. We wish to invert the relation - $\alpha\left(t\right)$, that is, what is the position variance given the temperature. The inversion has 3 distinct solutions, however imposing physical conditions on $\alpha$, that $\alpha\ge0$ and $\lim_{t\to0}\alpha\left(t\right)=0$ (or at least finite limit value) we obtain a unique solution which has non-zero imaginary part for $t\approx0.384$, maximal obtainable temperature for our constraints.
 Calculation of the equipartiton parameter yields -
\begin{align*}
\ex{p^{2}} & =\sigma_{p}^{2}\\
 & =mk_{B}T_{atoms}\\
 & =mtA\\
\ex{x^{2}} & =\alpha\left(t\right)\sigma_{t}^{2}\\
\chi_{H} & =\sqrt{\f{\sigma_{t}^{2}\ex{p^{2}}}{mA\ex{x^{2}}}}=\sqrt{\f t{\alpha\left(t\right)}}
\end{align*}
Fig.~\ref{fig:Analytical-calculation-of} shows how the equilibrium value of $\chi_{H}$ decreases as the temperature (normalized to trap depth) increases. This result supports the experimentally measured deviation from unity of the equipartiton parameter for a thermal state.
After confirming that the distribution reached its steady state we calculate the harmonic equipartiton parameter $\chi_{H}$ for different settings and $D$ values (Fig.~\ref{fig:chi_vs_D}). Within the simulation accuracy, for Brownian diffusion in harmonic trap (blue line) the equipartiton parameter remains unity for all $D$ values. Anomalous diffusion in harmonic trap (red line) shows clear deviation from unity of $\chi_{H}$, in agreement with~\cite{Dechant2016}. An important observation is the decline of $\chi_{H}$ from unity for Brownian diffusion in Gaussian trap (yellow line), in consistency with the analytical result of Fig.~\ref{fig:Analytical-calculation-of}. This can be explained by the fact $\chi_{G}$, and not $\chi_{H}$, is the correct parameter to quantify the equipartiton theorem in Gaussian trap. The effect is solely due to the anharmonic nature of the Gaussian potential. The deviation from unity increases as the spatial distribution of the steady state increases and samples more of the anharmonic nature of the confining potential. This claim can be supported by Fig.~\ref{fig:position-and-momentum_RMS_vs_D} showing that both
for Brownian diffusion and anomalous diffusion the spatial extension of the distribution grows monotonically with $D$. Finally we emphasize the combined effect of anomalous diffusion and anharmonic (Gaussian) trap (purple line). The data shows clear deviation of $\chi_{H}$ from unity, larger than the separate deviations of anomalous diffusion in harmonic trap or Brownian diffusion in Gaussian trap. We conclude that the effects of anomalous dynamics and anharmonicity cause additive deviations of $\chi_{H}$ from unity. For comparison we plot the Gaussian equipartiton parameter $\chi_{G}$ for Brownian diffusion and anomalous diffusion, and observe that for Brownian diffusion in a Gaussian trap $\chi_{G}$ remains unity, in contrasts to $\chi_{H}$. For Anomalous diffusion $\chi_{G}$ deviates from unity to smaller values, confirming the dependence of the equipartiton parameter upon the underlying mechanism of the diffusion process.
Next we compare the spread of the position and momentum distribution functions at the steady state, quantified by the RMS values (Fig.~\ref{fig:position-and-momentum_RMS_vs_D}). For both cases anomalous diffusion causes greater spread of the distribution, in comparison to Brownian diffusion.
In the discussion of Fig.~\ref{fig:chi_vs_D} we noticed that for the Brownian diffusion in Gaussian potential $\chi_{H}$ decreases with $D$, as the distribution samples more of the anharmonic potential. At this point we must verify that the extended deviation of $\chi_{H}$ for anomalous diffusion in comparison to Brownian diffusion in Gaussian trap is due to anomalous dynamics and not an artifact of the extended spatial distribution. To verify this point we plot (Fig.~\ref{fig:chi_vs_sigma_x}) the $\chi_{H}$ values as function of the RMS position (normalized by trap width). For a given RMS value of the position distribution there is a clear difference between anomalous diffusion and Brownian diffusion, attributed mainly by the different diffusion mechanisms. As an example we compare the difference in deviations of the harmonic equipartiton parameter in Gaussian trap for $D=0.1$. Switching from Brownian to anomalous diffusion causes a deviation of the equipartiton parameter $\Delta\chi_{H}$ in two different manners; a deviation that originates in extended spatial distribution and greater sampling of the anharmonic potential $\Delta_{1}$, and the deviation that originates from the anomalous diffusion mechanism $\Delta_{2}$. In a harmonic potential the deviation of the equipartiton parameter from unity is only due to the anomalous diffusion mechanism. We conclude that even though the Gaussian potential causes a deviation of the harmonic equipartiton parameter from unity, the anomalous diffusion mechanism causes a clear additional deviation of the harmonic equipartiton parameter from unity. Both deviations correspond to negative deviations of the equipartiton from unity, confirming our claim of additivity of the effects of anharmonic Gaussian potential and anomalous diffusion mechanism. 

\subsubsection*{Logarithmic perturbation to harmonic trap}

\begin{figure}[H]
\begin{centering}
\includegraphics[width=0.8\textwidth]{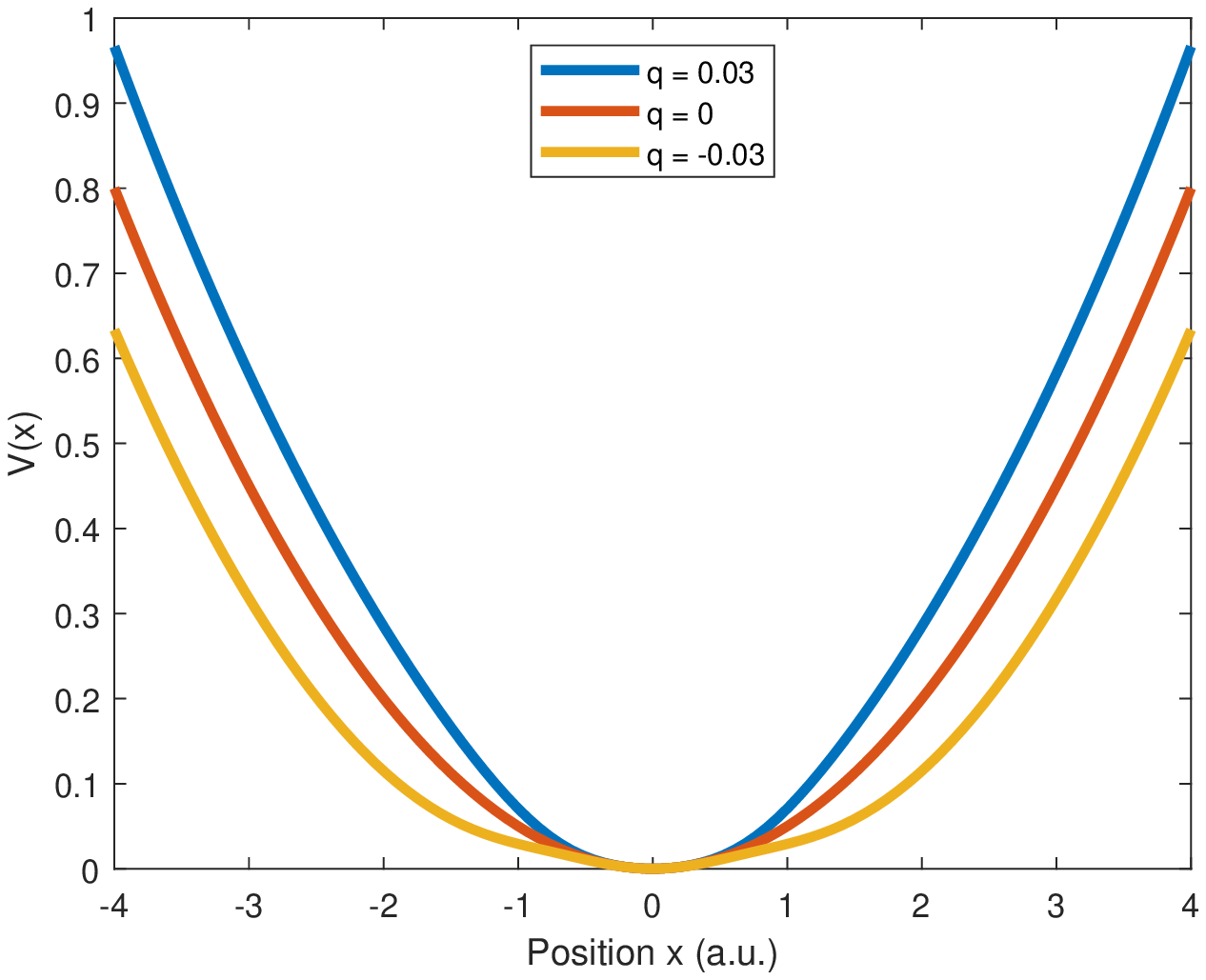}
\par\end{centering}
\caption{Harmonic potential with small logarithmic perturbation.\label{fig:Harmonic-potential-with-log-corr}}
\end{figure}

\begin{figure}[H]
\begin{centering}
\includegraphics[width=0.8\textwidth]{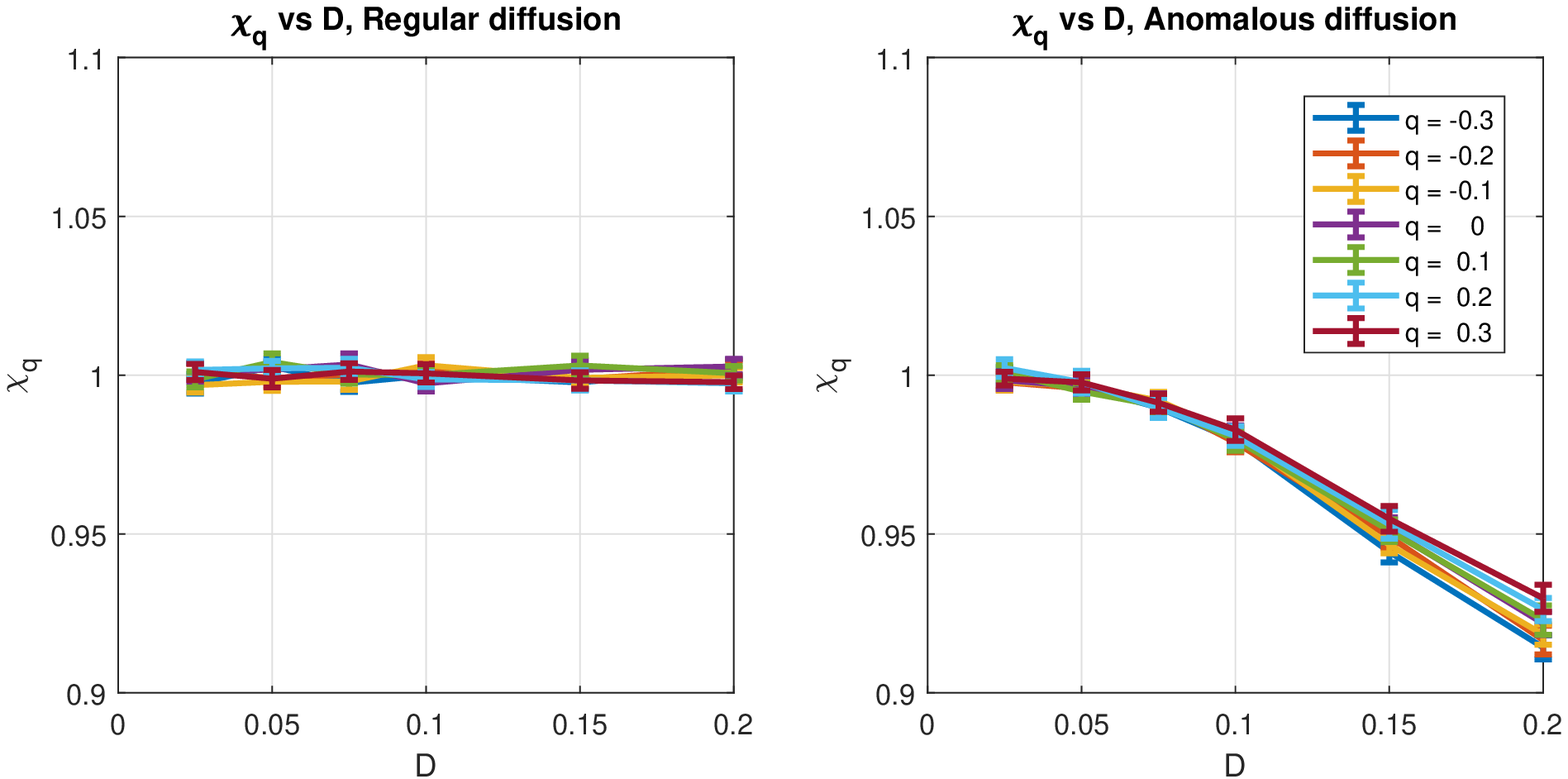}
\par\end{centering}
\caption{$\chi_{q}$ vs $D$\label{fig:chiq_vs_D}}
\end{figure}

\begin{figure}[H]
\begin{centering}
\includegraphics[width=0.8\textwidth]{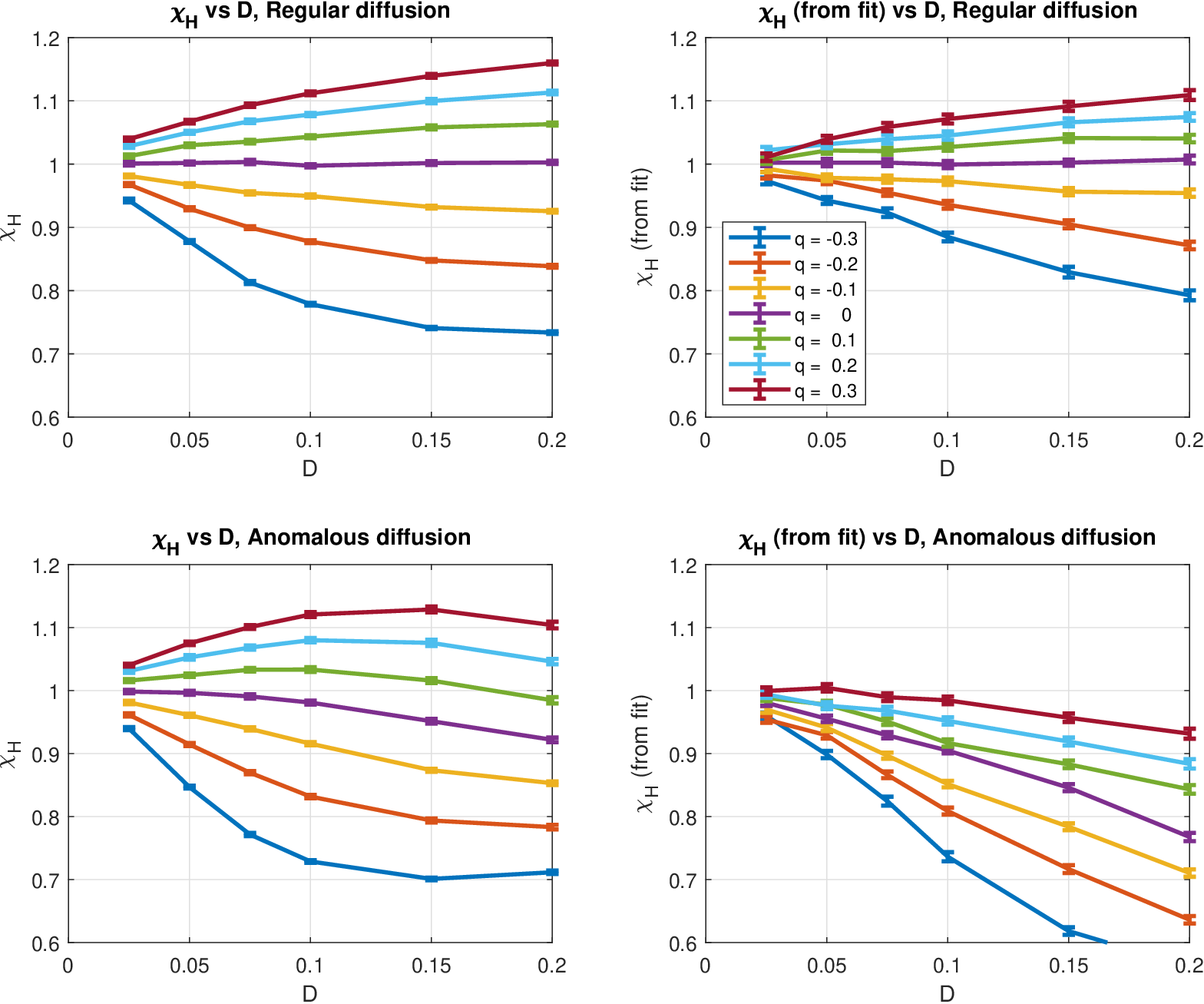}
\par\end{centering}
\caption{$\chi_{H}$ vs $D$\label{fig:chiH_vs_D}}
\end{figure}

\begin{figure}[H]
\begin{centering}
\includegraphics[width=0.8\textwidth]{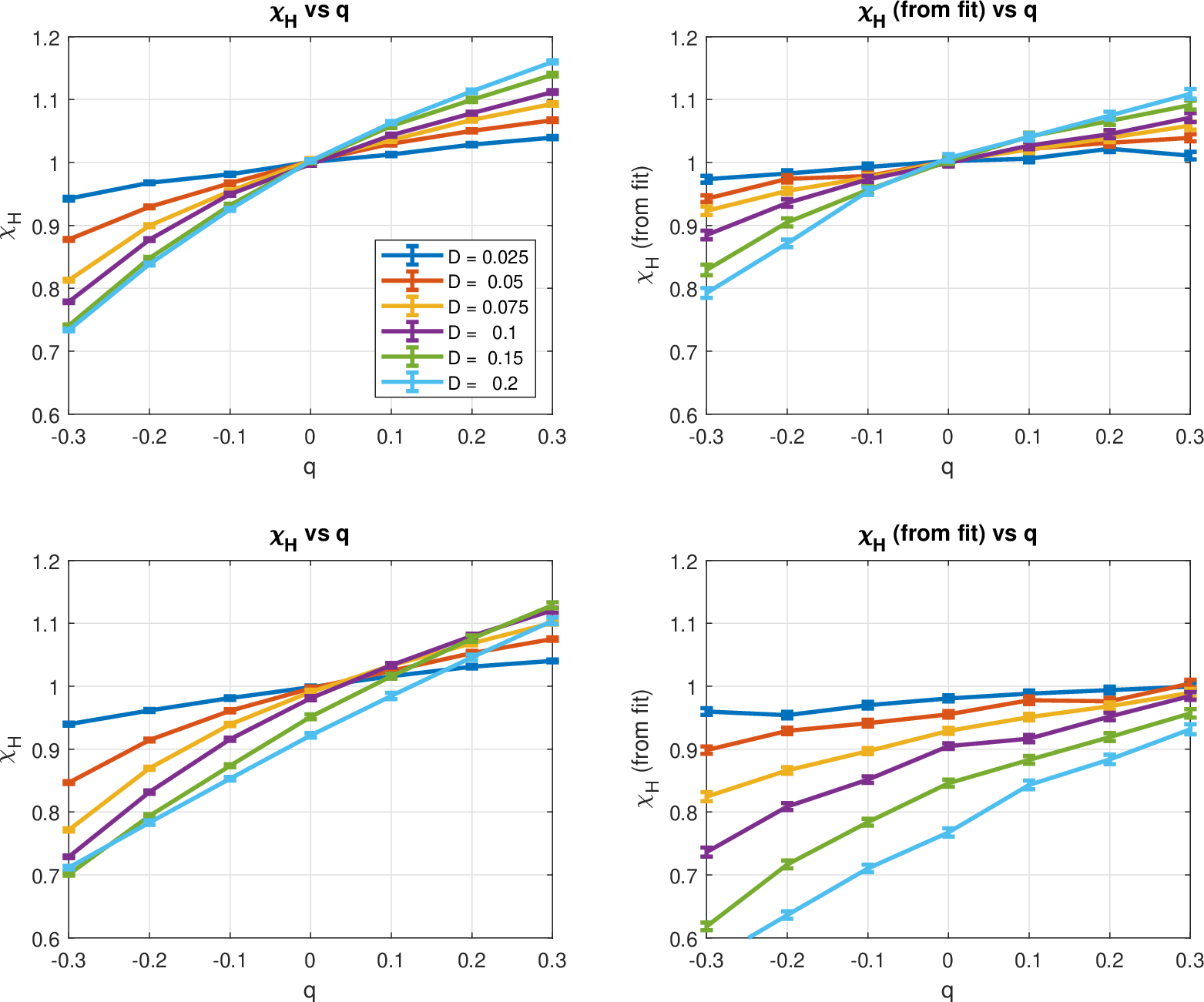}
\par\end{centering}
\caption{$\chi_{H}$ vs $q$ \label{fig:chiH_vs_q}}
\end{figure}

\begin{figure}[H]
\begin{centering}
\includegraphics[width=0.9\textwidth]{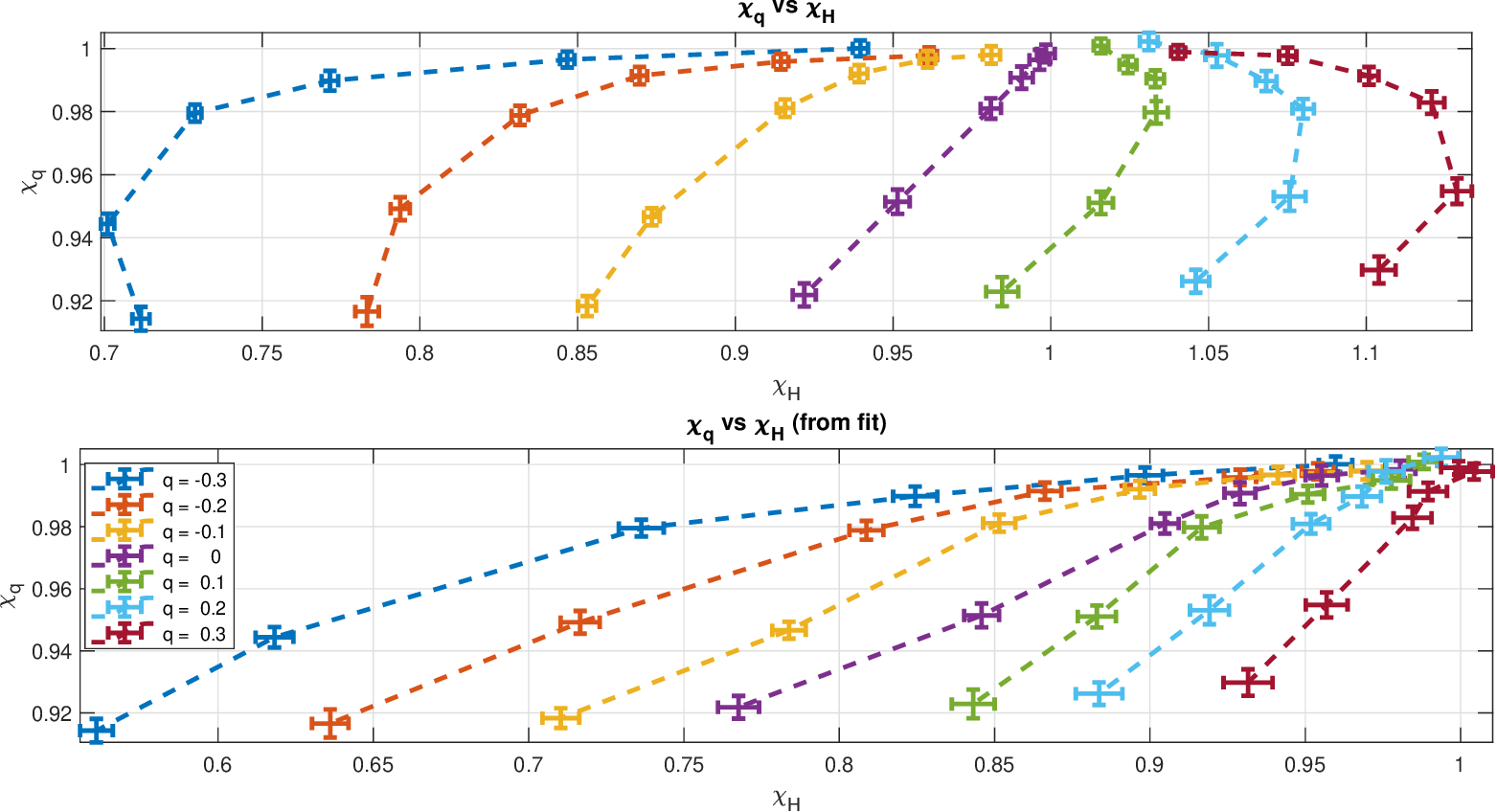}
\par\end{centering}
\caption{$\chi_{q}$ vs $\chi_{H}$\label{fig:chiH_vs_chiq}}
\end{figure}

The Gaussian potential presented in previous section is close to the actual physical confining potential, however the lack of stable steady states and a possibility to tune the \emph{sign}\footnote{By sign of anharmonicity we mean to the sign of the first non-vanishing term of the Taylor expansion of the potential, following the quadratic term, which essentially sets if the potential will be ``closing'' faster or slower than just the harmonic term} of the anharmonicity in it for our investigations of anharmoniciy effects lead us to simplified confining potential of harmonic potential and small logarithmic correction. We did not use the simplistic ``$x^{4}$'' correction since it has no steady state for negative prefactors either. The potential that was hence considered (Fig.~\ref{fig:Harmonic-potential-with-log-corr}) is, in dimensionless units -
\begin{align*}
V\left(x\right) & =\f 12\Omega x^{2}+q\log\left(1+x^{4}\right)\\
 & \approx\f 12\Omega x^{2}+qx^{4}+\O\left(x^{8}\right)
\end{align*}
The appropriate equipartiton parameter for this potential is - 
\[
\chi_{q}=\sqrt{\f{\ex{p^{2}}}{\ex{x^{2}+\f{4qx^{4}}{1+x^{4}}}}}
\]
Fig.~\ref{fig:chiq_vs_D} shows how $\chi_{q}$ changes for regular and anomalous diffusion as $D$ is increased.
As expected, for normal diffusion the parameter $\chi_{q}$ is independent of $D$ and is unity within numerical accuracy. For anomalous diffusion all simulated values of $q$ show larger deviations from unity as we increase $D$. Interestingly, all the deviations of $\chi_{q}$ are towards lower values, and the effect of $q$, which measures the amount of anharmonicity of the potential, is small in the specified regime. Notice however that $\chi_{q}$ parameter is considerably complex in comparison to the harmonic equipartiton parameter $\chi_{H}.$ The effect of anharmonic potential on $\chi_{H}$ is calculated using two methods: first by direct calculation of the desired moments from the phase space density distribution (PSD), and second using the marginal distribution function of position and momentum, fitting them to Gaussian functions and using the fitted Gaussians as the distribution with respect to which the expectation values were taken (Fig.~\ref{fig:chiH_vs_D}). In contrast to $\chi_{q}$, the values of $\chi_{H}$, if calculated directly from the PSD, can be both positive or negative. Another observation is the seemingly non-monotonic nature of $\chi_{H}$ vs $D$ for several $q$ values, if calculated directly by PSD. The use of Gaussian fits, however, seems to ``monotonise'' the dependence of $\chi_{H}$ on $D$ in this regime of $D$ values. Furthermore, for anomalous diffusion all the values obtained by Guassian fits are negative and decrease for larger $D$ values.
Next we investigated the dependence of $\chi_{H}$ on the anharmonicity parameter $q$ (Fig.~\ref{fig:chiH_vs_q}). For normal diffusion with specified $q$ values, both analysis methods result in qualitative agreement, as the sign of $\chi_{H}$ and of $q$ are the same. For normal diffusion all the lines cross the $\chi_{H}=1$ value at $q=0$, while for anomalous diffusion each line crosses at different $q$ value. In all cases $\chi_{H}$ increases with $q$. In most cases for a given value of $\left|q\right|$ the negative $q$ will have larger deviation in $\chi_{H}$ from unity than the positive one. That means that for cold enough atoms in a potential with negative anharmonicity, such as inverted Gaussian, the deviations of $\chi_{H}$ from unity are more pronounced than for appropriate potentials with positive anharmonicity. Moreover, the amount of negative anharmonicity of Gaussian traps can be easily tuned in order to increase the sensitivity of $\chi_{H}$ to Sisyphus lattice depth, manifested by $D$.
Lastly, Fig.~\ref{fig:chiH_vs_chiq} shows the ``calibration curve'' of $\chi_{q}$ vs $\chi_{H}$, that is, what is the appropriate $\chi_{q}$ equipartiton parameter for the measured $\chi_{H}$. A calculation of $\chi_{H}$ directly by PSD results in multi-valued function of $\chi_{q}\left(\chi_{H}\right)$ and is of little use. The calculation of $\chi_{H}$ from Gaussian fits however results in a monotonic function of $\chi_{q}\left(\chi_{H}\right)$ with favorable sensitivity $\f{\delta\chi_{H}}{\delta\chi_{q}}\gtrsim1$. That is, deviations of the real equipartiton parameter $\chi_{q}$ will cause even larger deviations of the $\chi_{H}$ parameter. $\chi_{H}$ that is calculated from Gaussian fits of position and momentum distributions is therefore a fair estimator of deviation of the correct equipartiton parameter for small anharmonicity of either sign. For negative anharmonicity ($q<0$) these $\chi_{H}$values have increased sensitivity to deviations of the correct equipartiton parameter.

\subsection*{Inhomogeneous local energy distributions in harmonic potential}

\begin{figure}[H]
\begin{centering}
\includegraphics[width=0.8\textwidth]{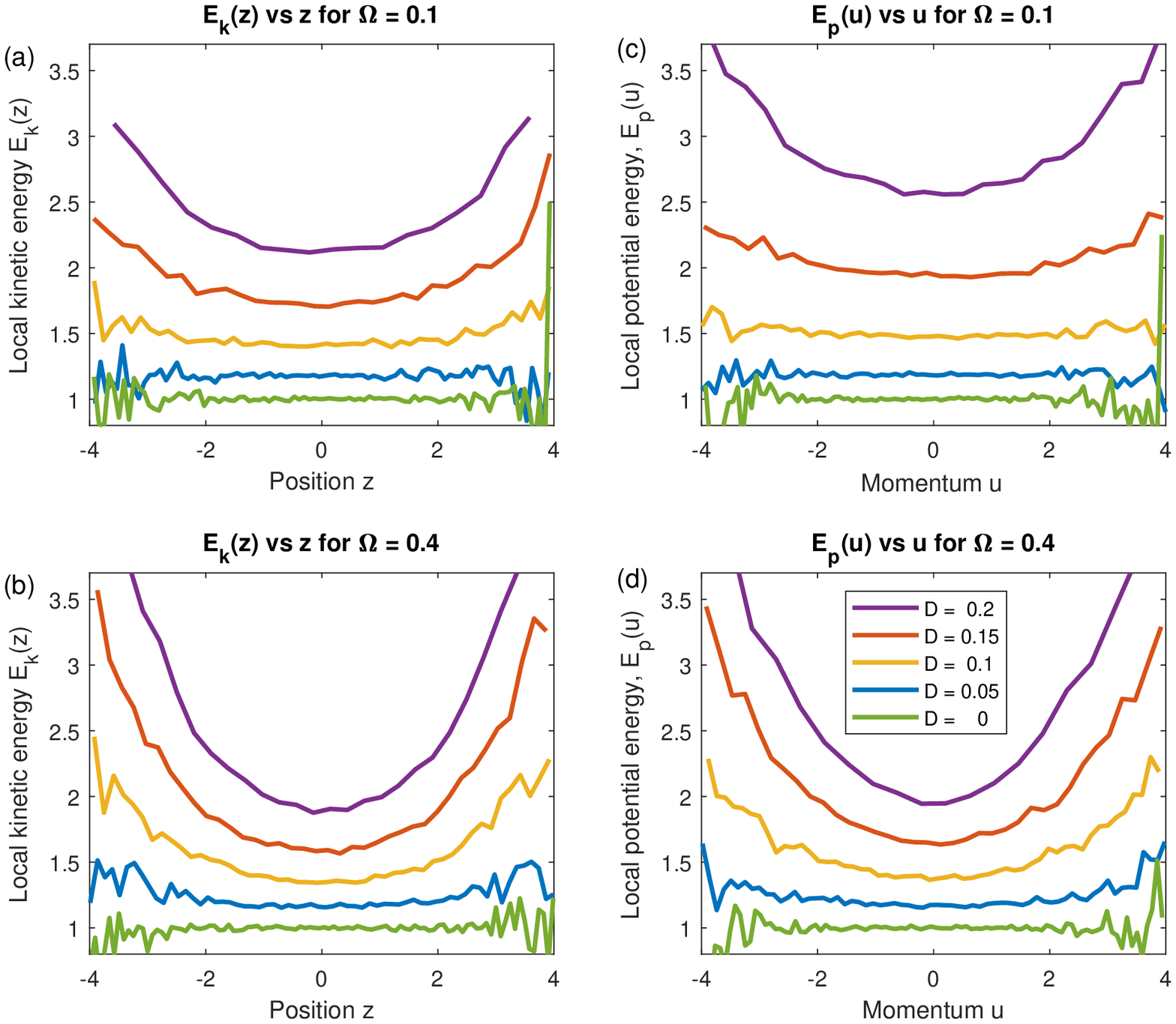}
\par\end{centering}
\caption{Simulation results of LED dependence on $z$ and $u$. (a)-(b) Mean
kinetic energy as function of position for $\Omega=0.1$ and $\Omega=0.4$,
respectively. (c)-(d) Mean potential energy as function of momentum
for $\Omega=0.1$ and $\Omega=0.4$, respectively.\label{fig:Temp_vs_z_u_Omega_parabols}}
\end{figure}

\begin{figure}[H]
\begin{centering}
\includegraphics[width=0.8\textwidth]{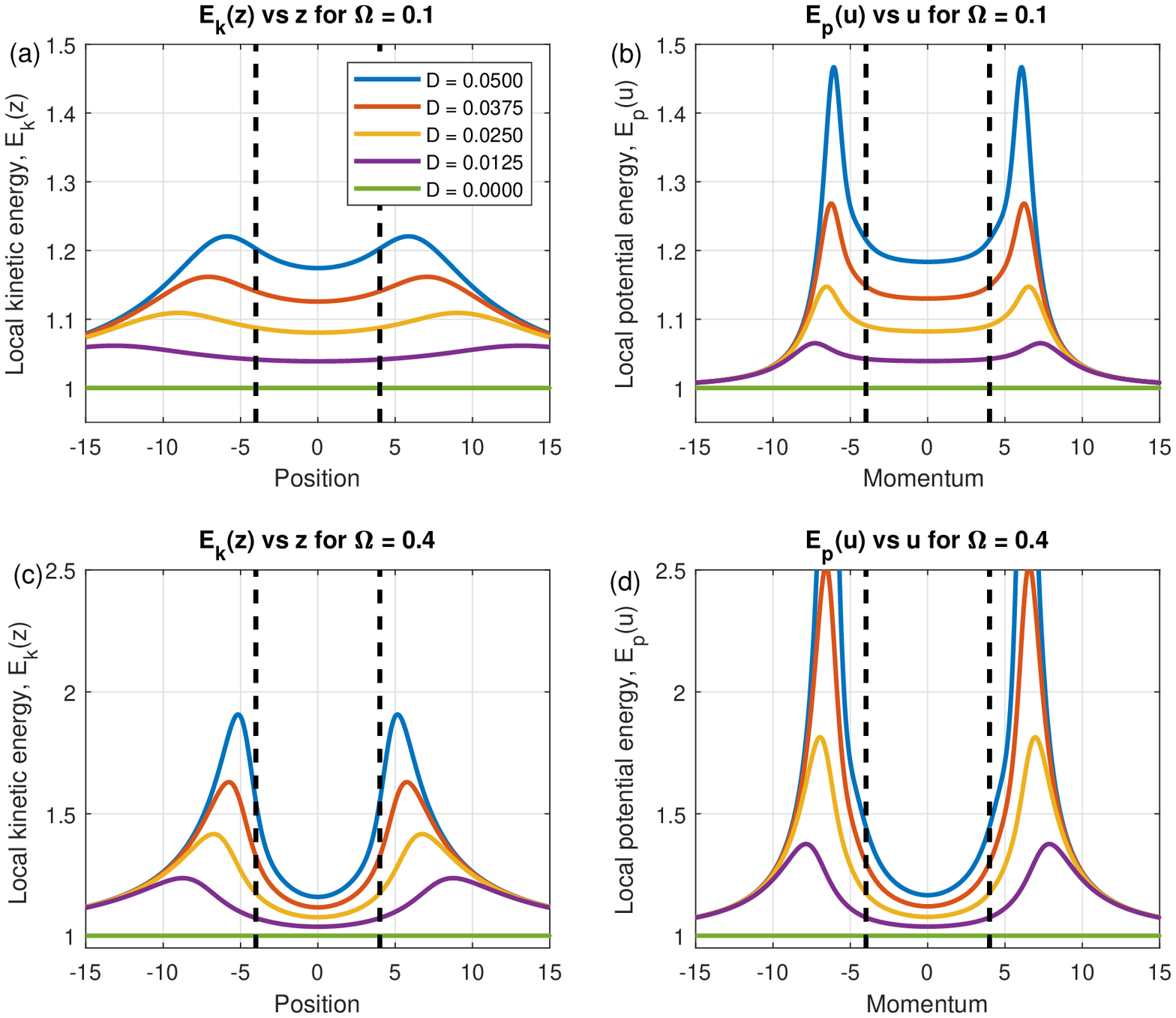}
\par\end{centering}
\caption{Analytical results of kinetic and potential energies dependence on
$z$ and $u$. (a)-(b) Mean kinetic energy as function of position
for $\Omega=0.1$ and $\Omega=0.4$, respectively. (c)-(d) Mean potential
energy as function of momentum for $\Omega=0.1$ and $\Omega=0.4$,
respectively.\label{fig:Analytical-results-of}}
\end{figure}

\begin{figure}[H]
\begin{centering}
\includegraphics[width=0.6\textwidth]{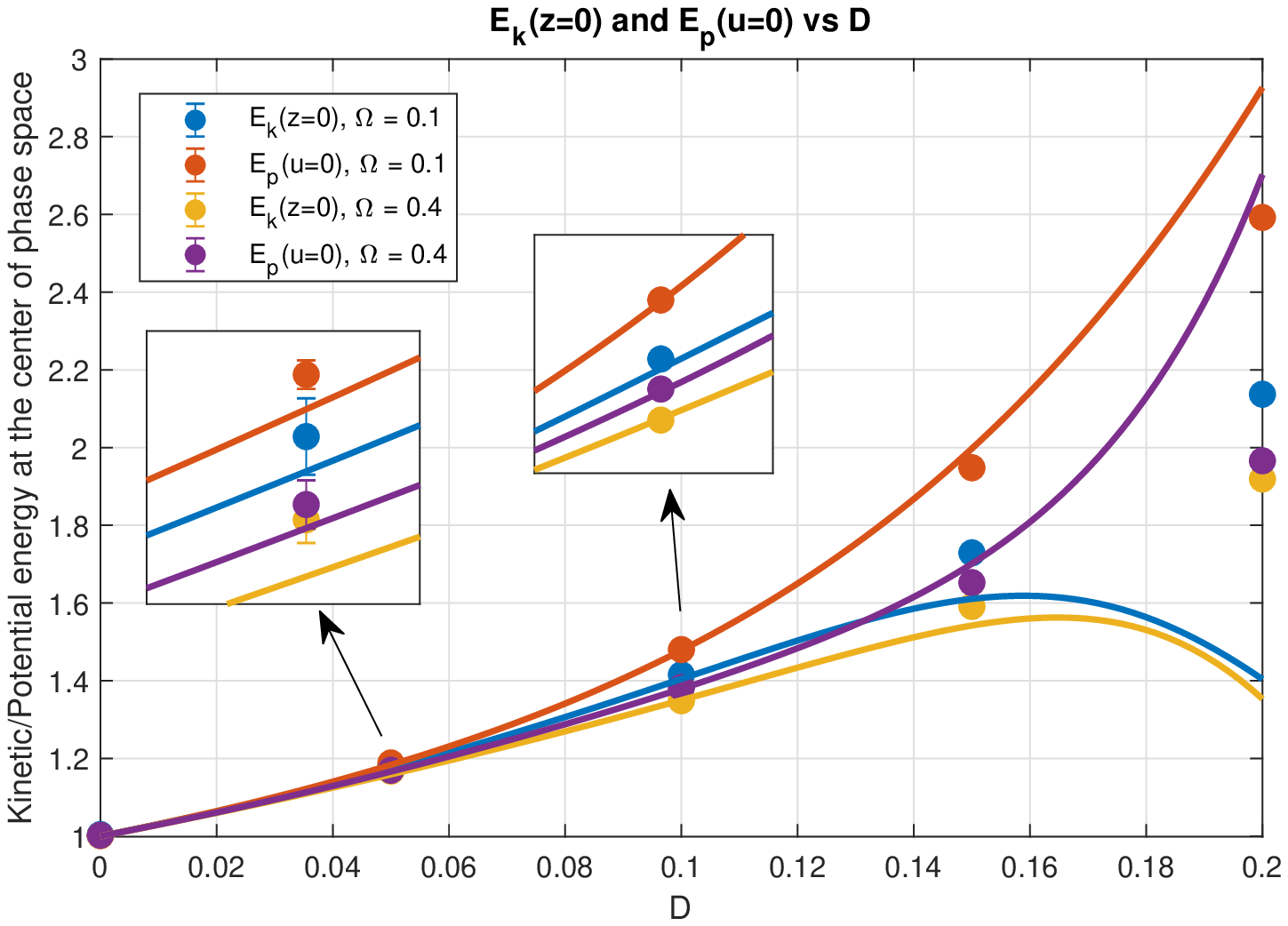}
\par\end{centering}
\caption{The values of position and momentum dependent temperatures at the
center of the distribution as function of $D$ for $\Omega=0.1$ and
$\Omega=0.4$.\label{fig:T0}}
\end{figure}

\begin{figure}[H]
\begin{centering}
\includegraphics[width=0.6\textwidth]{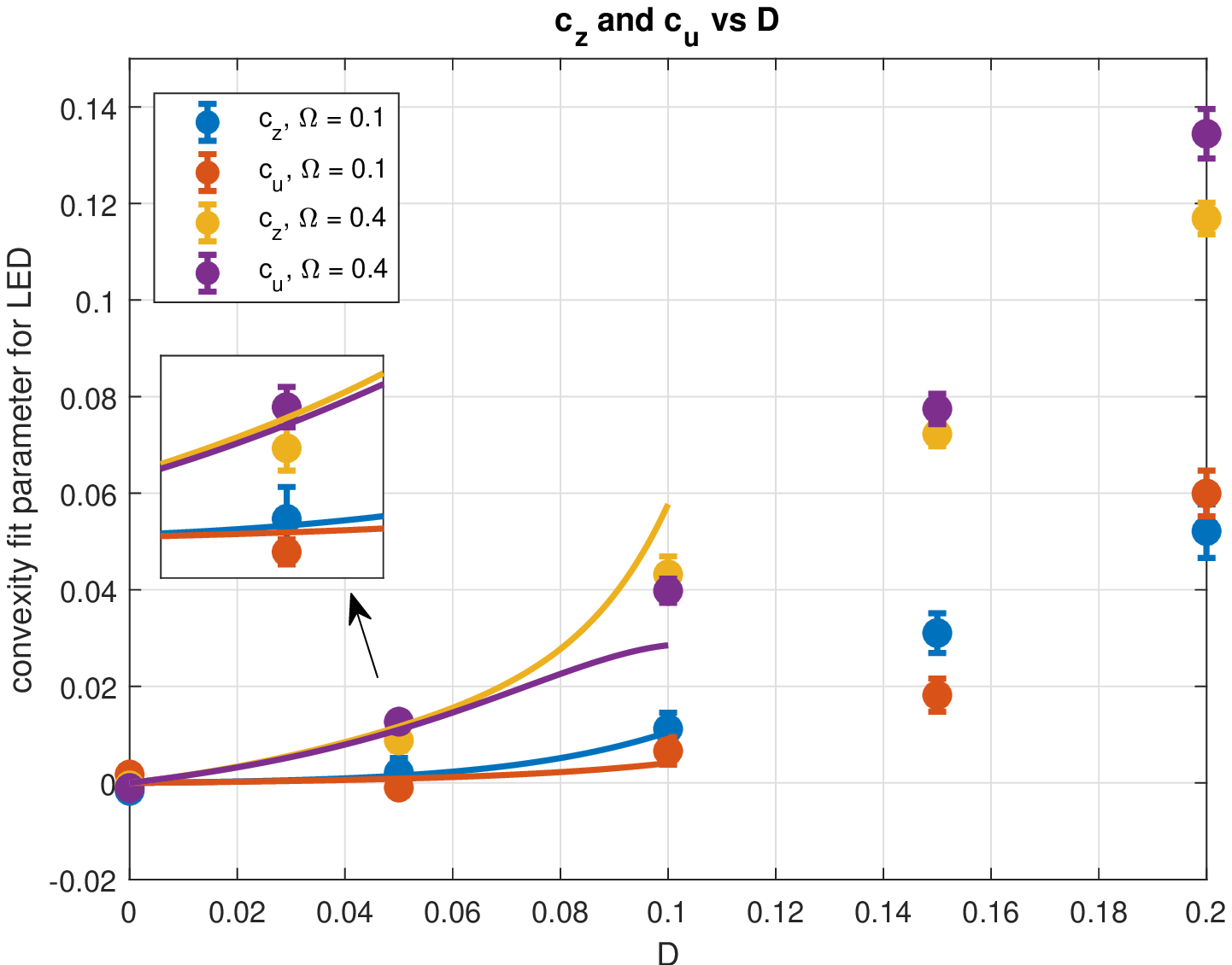}
\par\end{centering}
\caption{The convexity $c$ values (defined in the text) of position and momentum dependent temperatures at the center of the phase space distribution as function of $D$ for $\Omega=0.1$ and $\Omega=0.4$.\label{fig:pvalue}}
\end{figure}

The local kinetic and potential energy distributions (LED) were calculated by the following definitions -
\begin{align}
E_{k}\left(z\right) & :=\f{\int_{-\inf}^{\inf}u^{2}\rho\left(z,u\right)\dif u}{\int_{-\inf}^{\inf}\rho\left(z,u\right)\dif u}\label{eq:Ek}\\
E_{p}\left(u\right) & :=\f{\int_{-\inf}^{\inf}z^{2}\rho\left(z,u\right)\dif z}{\int_{-\inf}^{\inf}\rho\left(z,u\right)\dif z}\label{eq:Ep}
\end{align}
with $\rho\left(z,u\right)$ being the phase space distribution function at the steady state, obtained by a Monte-Carlo simulation or the perturbative phase space calculation for anomalous diffusion. The meaning of \emph{local} quantities above is that they are coordinate dependent, in contrast to equilibrium where the quantities are coordinate independent (and thus homogeneous through the system). We calculated the LED, defined by Eq.~\ref{eq:Ek}-\ref{eq:Ep} for various $D$ parameters and two $\Omega$ values in the over-damped limit. The simulation results are presented in Figure~\ref{fig:Temp_vs_z_u_Omega_parabols} and the analytical results in Figure~\ref{fig:Analytical-results-of}. A different range of $D$ parameter was used for simulation and analytical methods, for reasons that will be explained later. The black dashed lines in Figure~\ref{fig:Analytical-results-of} marks, for comparison, the horizontal limits of the presented simulation results, $z,u\in\left(-4,4\right)$ in Figure~\ref{fig:Temp_vs_z_u_Omega_parabols}. For all simulated cases the range of $z,u\in\left(-4,4\right)$ contains about $99\%$ of the particles. The edges of the presented simulation data were chosen to exclude the noisy tails, where the SNR of the available simulation data is low.
For Brownian diffusion (green line) we notice a flat distribution\footnote{For all simulation cases the low signal-to-noise ratio at the tails of the distributions causes the fluctuations in energy values for large $\left|z\right|$ or $\left|u\right|$ values.} for both analytical and simulation results, in consistency with the separability property of the Boltzmann-Gibbs distribution. However, as we cross from Brownian diffusion ($D=0$) to the anomalous regime ($D>0$), we notice a clear and monotonic emergence of non-uniformity of all the presented LEDs. We notice that for certain settings the LED values can have as much as 70\% variation in comparison to the values at the center of the distribution. The analytical results show emergence of maxima for all LED profiles, and later asymptotic decay to equilibrium temperature for larger coordinate values. This phenomena was not captured by the simulations, and may be an artifact of slow convergence rate of the of the phase space expansion series. It has been previously stressed~\cite{Dechant2016} that the phase space expansion is only valid for the central part of the phase space (small $\left|z\right|$ and $\left|u\right|$ values) and for small $D$. The leading order behavior of LED in this regime shows cubic coordinate dependence, so phenomenologically, 
\begin{equation}
E_{x}=E_{x}\left(0\right)+c_{x}\cdot w^{2},\qquad w\in\left\{ z,u\right\} \label{eq:parabola_fit}
\end{equation}
with $E_{x}\left(0\right)$ and $c_{x}$ being fit parameters of the appropriate kinetic or potential energy distributions. Figures~\ref{fig:T0}-\ref{fig:pvalue} present these fit parameters for all the simulation data presented in Fig.~\ref{fig:Temp_vs_z_u_Omega_parabols}. The solid lines represent the theoretical predictions by using phase space expansion to order $D^{7}$ with appropriate $D$ and $\Omega$ values. The markers represent the parameters extracted from simulation data and fitted to centered parabola, described by Eq.~\ref{eq:parabola_fit}.
There is a general agreement of simulation and analytical results up to $D\approx0.1$. The qualitative and quantitative comparison of central energy values, $E_{x}\left(0\right)$, presented in Fig.~\ref{fig:T0}, are in good agreement in this regime. The comparison for LED convexity parameter $c$ at the center is presented in Fig.~\ref{fig:pvalue} and shows qualitative, but not quantitative agreement in the specified $D$ regime. The calculated analytical results of $c$ parameter for $D>0.1$ were omitted, due to instability of the analytical result in this regime, far from the validity regime of the phase space expansion.
Examining Figs.~\ref{fig:T0}-\ref{fig:pvalue} and identifying the mean local energies as local equilibrium temperatures we notice several general effects. The first is the effect of $\Omega$ on central convexity $c$, which increases for larger $D$ values. However for $D\to0$ all the results converge to a single value. This is interesting especially in context of total temperature, or temperature at the center of the distribution in contrast to Brownian diffusion. For systems in contact with thermal baths, the temperature is defined only by the bath properties, and the micro structure of the system has no effect. Here we notice how the micro structure of the system
affects the temperature profile. Examining the central temperatures behavior in Fig.~\ref{fig:T0} we notice that even for the same $D$ value the central position temperatures $E_{k}\left(0\right)$ and central momentum temperatures $E_{p}\left(0\right)$ are sufficiently different. This claim is supported both by analytical result and the errorbars of the simulation results, which distinguish the different values. We notice that in general for a given $D$ value the central temperatures are higher for smaller $\Omega$ values. For fixed $\Omega$ and $D$, $E_{p}\left(0\right)$ values are higher than $E_{k}\left(0\right)$, meaning that the average potential energy of particles at rest is greater than average kinetic energy of particles located at the minimum of the potential. All central temperatures obtained by simulation increase with $D$. Examining the parabola convexity parameter $c$ in Fig.~\ref{fig:pvalue} we notice that for fixed $\Omega$ value $E_{k}\left(z\right)$ and $E_{p}\left(u\right)$ have similar convexity. On the other hand we see that $c$ increases substantially with $\Omega$. The central convexity of all temperature profiles obtained by simulation increase with $D$.

\begin{figure}[H]
\begin{centering}
\includegraphics[width=0.8\textwidth]{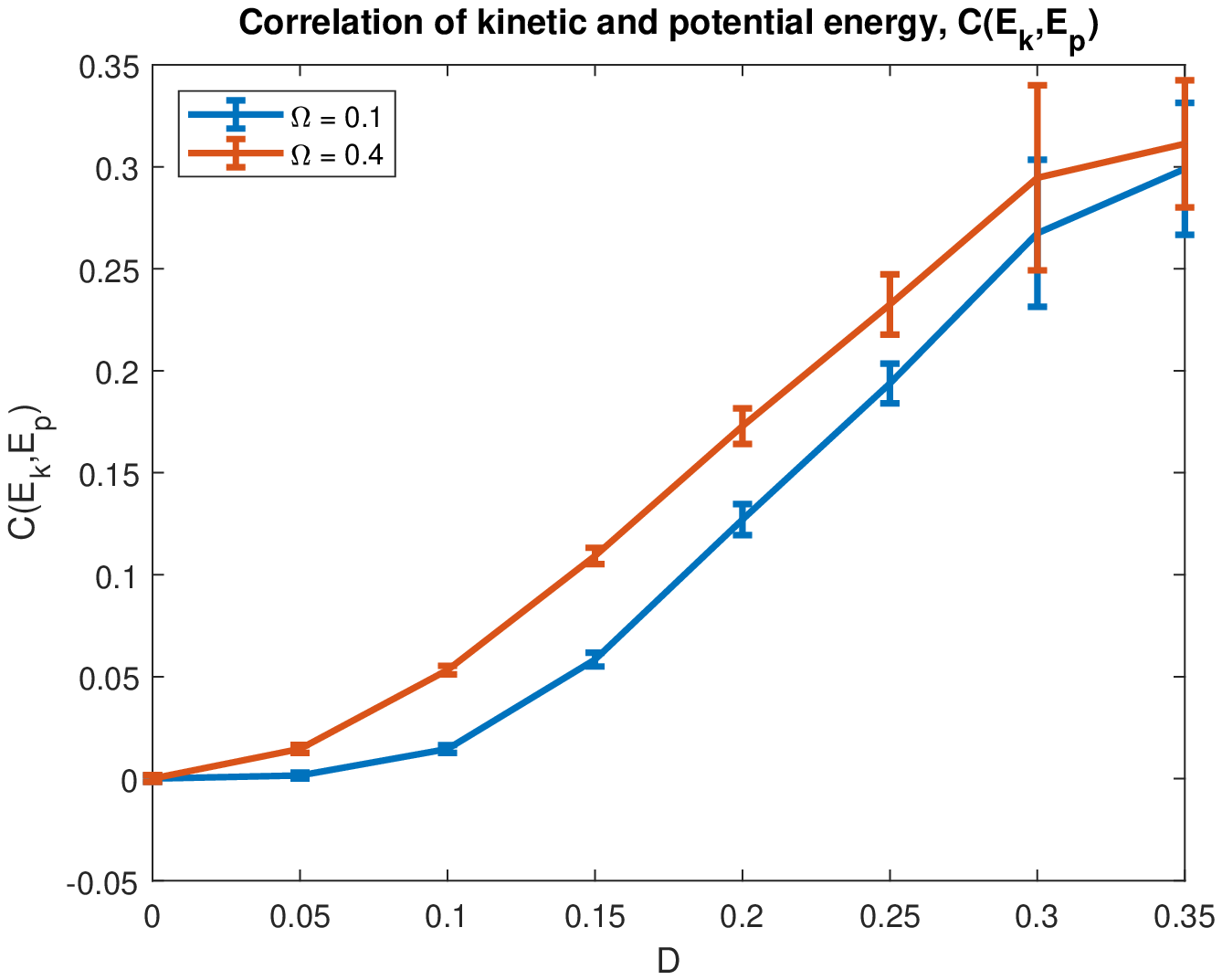}
\par\end{centering}
\caption{Correlation of potential end kinetic energies as function of the dimensionless diffusion parameter for two distinct trap frequencies $\Omega$.\label{fig:Energy-correlations}}
\end{figure}

Lastly we calculate the kinetic-potential energy correlations of the simulation results using the following relation (Fig.~\ref{fig:Energy-correlations})
\[
C\left(E_{k},E_{p}\right)=\f{\ex{z^{2}u^{2}}-\ex{z^{2}}\ex{u^{2}}}{\sqrt{\ex{z^{4}}-\ex{z^{2}}^{2}}\sqrt{\ex{u^{4}}-\ex{u^{2}}^{2}}}
\]

As $D$ increases and with it the anomality of the dynamics, the kinetic and potential energies become substantially correlated. We also notice the dependence of the correlation value on system parameter $\Omega$, consolidating the dependence of non-equilibrium steady state on system parameters, in opposition to equilibrium state ($D=0$) where the two curves converge.


\pagebreak
\bibliographystyle{apsrev4-1}
\bibliography{equipartition_supplementary}